\documentclass[12pt]{article}

\usepackage[dvips]{graphicx}
\usepackage{epsfig}
\usepackage{amsmath,amsfonts,amssymb,amsthm}
\usepackage{mathrsfs,mathtools}
\usepackage{verbatim}
\usepackage{psfrag}
\usepackage{bm}
\usepackage{bbm}
\usepackage[utf8]{inputenc}
\usepackage[square,comma,sort&compress,numbers]{natbib}
\usepackage[dvipsnames]{xcolor}
\usepackage{slashed}
\usepackage{upgreek}
\usepackage[normalem]{ulem}
\usepackage{enumitem}
\usepackage{float}
\usepackage[T1]{fontenc}
\usepackage{soul}
\usepackage{pgfplots}
\usepackage{extarrows}
\pgfplotsset{compat=1.17}
\usetikzlibrary{external,decorations.pathmorphing}
\usepackage{cases}
\usepackage{listings}
\usepackage{cancel}

\lstdefinestyle{mystyle}{
    basicstyle=\ttfamily\footnotesize,
    breakatwhitespace=false,         
    breaklines=true,                 
    captionpos=b,                    
    keepspaces=true,                 
    numbers=left,                    
    numbersep=5pt,                  
    showspaces=false,                
    showstringspaces=false,
    showtabs=false,                  
    tabsize=2
}
\usepackage{lmodern}
\usepackage{tikz}

\usepackage[T1]{fontenc}

\tikzset{
  external/only named=true,
  thick/.style={line width=.5pt},
  approximation/.style={line width=1.2pt},
  numerics/.style={black, dotted, line width=.8pt},
  amplitude/.style={dashed},
  estimate/.style={dashed, line width=.8pt},
  normal plot/.style={line width=.8pt},
}

\tikzset{snake it/.style={decorate, decoration=snake}}

\usepackage[colorlinks=true,urlcolor=blue,anchorcolor=blue,citecolor=blue,filecolor=blue
,linkcolor=blue,menucolor=blue,linktocpage=true,pdfproducer=medialab,pdfa=true
]{hyperref}

\usepackage{epsf,epsfig}
\usepackage{graphics}
\usepackage{subcaption}

\newcommand\blfootnote[1]{%
  \begingroup
  \renewcommand\thefootnote{}\footnote{#1}%
  \addtocounter{footnote}{-1}%
  \endgroup
}

\def\d{\mathrm{d}}

\def\L{\mathcal{L}}

\def\O{\mathcal{O}}
\def\vec{\mathbf}

\def\e{\mathrm{e}}
\def\P{\mathcal{P}}

\def\M{\mathcal{M}}
\def\veck{\vec{k}}
\def\vecp{\vec{p}}

\newcommand{\overbar}[1]{\mkern 1.5mu\overline{\mkern-1.5mu#1\mkern-1.5mu}\mkern 1.5mu}

\usepackage[errorstop]{feynmp}
\newcommand{\lsim}
{\;\raisebox{-.3em}{$\stackrel{\displaystyle <}{\sim}$}\;}

\newcommand{\rom}[1]{\uppercase\expandafter{\romannumeral #1\relax}}

\begin{document}                                                                                                                                                                                                                                                                

\thispagestyle{empty}

\begin{flushright}
{
\small
KCL-PH-TH/2024-38
}
\end{flushright}

\vspace{-0.5cm}

\begin{center}
\Large\bf\boldmath
Non-thermal production of heavy vector dark matter from relativistic bubble walls
\unboldmath
\end{center}

\vspace{-0.2cm}

\begin{center}
Wen-Yuan Ai,$^{*1}$\blfootnote{$^*$~wenyuan.ai@kcl.ac.uk} 
Malcolm Fairbairn,$^{\dagger 1}$\blfootnote{$^\dagger$~malcolm.fairbairn@kcl.ac.uk} Ken Mimasu$^{\ddagger 2}$\blfootnote{$^\ddagger$~tevong.you@kcl.ac.uk} and Tevong You$^{\S 1}$\blfootnote{$^\S$~ken.mimasu@soton.ac.uk}  \\
\vskip0.4cm
{\it $^1$Theoretical Particle Physics and Cosmology, King’s College London,\\ Strand, London WC2R 2LS, United Kingdom} \\
\vskip0.4cm
{\it $^2$School of Physics and Astronomy, University of Southampton,\\
Highfield, Southampton S017 1BJ, United Kingdom} 
\vskip1.cm

\end{center}

\begin{abstract}
Heavy vector boson dark matter at the TeV scale or higher may be produced non-thermally in a first-order phase transition taking place at a lower energy scale. 
While the production of vector dark matter has previously been studied for bubble wall collisions, here we calculate production by bubble wall expansion in a plasma, which can be the dominant production mechanism. We compute the results numerically and provide an analytical fit for the vector dark matter density. The numerical fit is also validated for scalar dark matter production, obtaining results in agreement with past literature. We find that vector pair production leads to bubble wall friction with a novel boost factor scaling behaviour compared to transition radiation emission of a single vector. We conclude that TeV-scale WIMP vector dark matter can be efficiently produced non-thermally by first-order phase transitions in a wide region of parameter space where thermal freeze-out is inefficient. In this scenario, the phase transition scale is predicted to be in the sub-GeV to $\mathcal{O}(10)$ TeV range and could therefore be accessible to future gravitational wave detectors. 
\end{abstract}

\newpage

\hrule
\tableofcontents
\vskip .85cm
\hrule


\section{Introduction}
\label{sec:Intro}

Despite a substantial body of observational evidence for dark matter~\cite{Bertone:2016nfn, Cirelli:2024ssz}, our understanding of its particle nature and origins in the early universe remains extremely limited. Weakly interacting massive particles (WIMPs), and their production through the thermal freeze-out process, are one of the simplest and most promising candidates for dark matter and its production mechanism~\cite{Lee:1977ua,Srednicki:1988ce,Gondolo:1990dk,Griest:1990kh}. In this scenario, dark matter particles were once in thermal equilibrium with the Standard Model (SM) plasma but underwent freeze-out as their interactions with the plasma failed to keep pace with the expansion of the Universe. Since their relic density is fixed by the efficiency with which they annihilate away before freezing out, it typically increases with greater dark matter mass and diminishes with stronger interaction strength. 
In the freeze-out framework, the mass of dark matter particles is capped at an upper limit of about $100$ TeV due to unitarity constraints~\cite{Griest:1989wd,Baldes:2017gzw,Smirnov:2019ngs}, a threshold known as the Griest–Kamionkowski (GK) bound.

Alternative dark matter production mechanisms, that exhibit non-thermal characteristics, have been proposed. A notable example is the freeze-in mechanism~\cite{Hall:2009bx,Bernal:2017kxu}, where dark matter particles are generated from the primordial plasma starting from an initial vacuum density, never reaching thermal equilibrium due to their feeble interactions with the SM plasma. Such non-thermal mechanisms result in an inverse relationship between the relic abundance and interaction strength. The sheer scope of possibilities for the nature of dark matter, as well as its potential experimental and observational detection signatures, motivates exploring other production mechanisms and their resulting phenomenology. 

In recent years, several non-thermal dark matter production mechanisms have been proposed involving first-order phase transitions (FOPTs). For instance, heavy scalar dark matter production has been investigated during the rapid expansion of bubble walls~\cite{Azatov:2021ifm,Azatov:2022tii,Baldes:2022oev} and from dramatic wall collisions~\cite{Watkins:1991zt,Konstandin:2011ds} for the case of scalar, fermion, and vector dark matter~\cite{Falkowski:2012fb,Freese:2023fcr,Mansour:2023fwj,Shakya:2023kjf,Giudice:2024tcp}, or from filtering effects~\cite{Baker:2019ndr,Chway:2019kft,Marfatia:2020bcs,Borah:2025wzl}.\footnote{For other impacts on dark matter from FOPTs, see e.g. Refs.~\cite{Cohen:2008nb,Shelton:2010ta,Petraki:2011mv,Chung:2011hv,Chung:2011it,Baker:2016xzo,Baldes:2017rcu,Hambye:2018qjv,Bian:2018mkl,Heurtier:2019beu,Hall:2019rld,Baldes:2020kam,Elor:2021swj,Wong:2023qon,Gehrman:2023qjn}.} FOPTs are particularly intriguing because they can lead to interesting phenomenology such as the generation of the cosmic matter-antimatter asymmetry~\cite{Kuzmin:1985mm,Morrissey:2012db,Garbrecht:2018mrp,Cline:2020jre,Azatov:2021irb,Baldes:2021vyz,Huang:2022vkf,Chun:2023ezg}, production of Fermi and Q balls~\cite{Hong:2020est,Jiang:2024zrb}, magnetic fields~\cite{Vachaspati:2001nb,Ellis:2019tjf,Di:2020kbw,Balaji:2024rvo}, and the formation of primordial black holes~\cite{Kodama:1982sf,Hawking:1982ga,Garriga:2015fdk,Deng:2017uwc,Gross:2021qgx,Baker:2021nyl,Kawana:2021tde,Liu:2021svg,Jung:2021mku,Huang:2022him,Lewicki:2023ioy,Gouttenoire:2023naa,Lewicki:2024ghw,Flores:2024lng,Ai:2024cka}. In particular, FOPTs can generate a stochastic gravitational wave background (SGWB)~\cite{Witten:1984rs,Kosowsky:1991ua,Kosowsky:1992vn,Kamionkowski:1993fg,Huber:2008hg,Hindmarsh:2013xza} potentially detectable with current and future gravitational wave (GW) detectors~\cite{Grojean:2006bp,Caprini:2018mtu,Caprini:2019egz,LISACosmologyWorkingGroup:2022jok}.

The production of very heavy particles from a fast bubble wall (with the Lorentz factor of the wall velocity, $\gamma_w\gg 1$) is possible because $(i)$ the wall (assuming a planar wall expanding in the $z$-direction) breaks $z$-translation symmetry and thus the total $z$-momentum of the particles in a microscopic particle process is not necessarily conserved and $(ii)$ the energies of particles moving towards the wall are boosted by $\gamma_w$  compared to the nucleation temperature~\cite{Bodeker:2017cim,Azatov:2020ufh,Azatov:2021ifm,Ai:2023suz}. Particles much heavier than the phase transition scale, which would be Boltzmann suppressed right before the phase transition, can therefore be produced from bubble expansion, and indeed also from bubble collisions~\cite{Falkowski:2012fb,Giudice:2024tcp,Baldes:2024wuz}. 

In this work, we study the non-thermal production of heavy vector dark matter from the expansion of relativistic bubble walls. This has previously been considered for the case of bubble collisions; however, as we argue in Sec.~\ref{sec:BEvsBC}, bubble expansion in a thermal plasma can generically be the dominant production mechanism. Moreover, the production of vector bosons in FOPTs is of particular phenomenological relevance for bubble wall dynamics. It is well known that extra friction, linear in 
$\gamma_w$, is introduced by transition radiation, where a gauge boson is emitted by a particle as it passes through the bubble wall~\cite{Bodeker:2017cim,Gouttenoire:2021kjv}. This effect prevents electroweak walls from reaching the relativistic velocities needed for significant particle production. One may then ask whether the decay of an excitation of the scalar undergoing the FOPT into a pair of vector bosons, such as our dark matter candidate, leads to similar friction. In Sec.~\ref{sec:friction} we find that the pressure from vector pair production resulting from bubble expansion exhibits a novel quadratic dependence on $\gamma_w$ (which may arise from a $\gamma_w \log{(1+c\gamma_w)}$ scaling when $c \gamma_w > 1$) but with a small overall coefficient such that the bubble wall can still achieve highly relativistic velocities.  

We illustrate the potential impact of non-thermal bubble wall production of vector dark matter for the case of TeV-scale WIMPs whose annihilation cross-section is too large to generate the observed dark matter density through thermal freeze-out. As the temperature drops below the phase transition temperature of a dark sector scalar, a FOPT is triggered and TeV-scale vector dark matter coupled to the scalar is abundantly produced by relativistic bubble walls expanding through a plasma containing the scalar particle. The number density of this non-thermal population of vector bosons is then subsequently reduced by annihilations to the observed relic density. We find that the required phase transition temperature is likely to be in the sub-GeV to $\mathcal{O}(10)$ TeV range. This raises the intriguing possibility of correlating the mass and couplings of WIMP dark matter with a possible detection of stochastic GW signals at future GW observatories. 

This paper is organised as follows. In Sec.~\ref{sec:BEvsBC} we provide a rough criterion for when the fraction of vacuum energy going into the plasma is larger than the energy in bubble wall collisions. Sec.~\ref{sec:scalar} starts with the case of scalar dark matter production in order to validate our numerical calculation and methods with previously known results. We also provide convenient analytical fitting formulae to our numerical results to ease future phenomenological analyses. We then study the case of vector dark matter production in Sec.~\ref{sec:vectorproduction}, comparing its effect on bubble wall friction with the scalar dark matter and transition radiation case in Sec.~\ref{sec:friction}, and apply this production mechanism to TeV-scale WIMP vector dark matter in Sec.~\ref{sec:wash-out}. GW signals are discussed in Sec.~\ref{sec:GW} before some concluding remarks in Sec.~\ref{sec:Conc}.

{$\left. \right.$}

{\it Note added:} As this work was being completed, a preprint~\cite{Azatov:2024crd} appeared which also considers vector dark matter production from bubble expansion. Their interesting and complementary study focuses on interactions through higher-dimensional operators with heavier and more weakly-coupled dark matter closer to the freeze-in region of parameter space.

\section{Bubble expansion versus bubble collision}
\label{sec:BEvsBC}

When dark matter is coupled to the order-parameter scalar that undergoes a FOPT, it can be produced from both bubble expansion~\cite{Azatov:2021ifm} and bubble collisions~\cite{Falkowski:2012fb, Giudice:2024tcp}. Therefore, one may wonder which process dominates the dark matter production. This is a difficult question whose answer depends on the production efficiency of each process as well as the fractions of the vacuum energy transferred into kinetic energy of the bubble wall, $\kappa_{\rm wall}$, and the plasma, $\kappa_{\rm plasma}$. The production efficiencies require detailed studies of the microscopic particle processes and are therefore rather model-dependent. In this section, we shall briefly discuss the energy fractions $\kappa_{\rm wall}$ and $\kappa_{\rm plasma}$, and take the rough criterion that one process dominates over the other when its corresponding energy fraction is larger.

The first quantity we need is the terminal wall velocity $v_w$.
Usually, its determination is very complicated, requiring one to solve the Boltzmann equations for the particle distribution functions (which are integro-differential equations), the background scalar equation of motion, and the fluid equations for the hydrodynamics~\cite{Dine:1992wr,Liu:1992tn,Moore:1995ua,Moore:1995si}. In this work, we shall simply assume that after nucleation the bubble wall accelerates to a large terminal wall velocity $v_w$ very quickly, entering the so-called detonation regime, as the dark matter production mechanism under study relies on relativistic bubble walls. 

With the terminal wall velocity $v_w$, the averaged final radius of the bubbles at collision is~\cite{Hindmarsh:2020hop}
\begin{align}
    R_*= \frac{(8\pi)^{\frac{1}{3}}v_w}{\beta}\,,
\end{align}
where $\beta$ is the transition rate, a mass dimension one parameter related to bubble nucleation rates, that characterises the phase transition. The total vacuum energy released per bubble is 
\begin{align}
    E_{\rm vac}=\frac{4\pi R_*^3}{3}\Delta V\,,
\end{align}
where $\Delta V$ is the difference in the zero-temperature potential of the symmetric and broken phases, $\Delta V= V_{s}^{(T=0)}-V^{(T=0)}_{b}$. The averaged bubble kinetic energy is given by~\cite{Ai:2020vhx}
\begin{align}
    E_{\rm wall}=4\pi \sigma R_*^2 \gamma_w\,,
\end{align}
where $\gamma_w\equiv 1/\sqrt{1-v_w^2}$ is the Lorentz factor of the {\it terminal} wall velocity and $\sigma$ is the wall surface tension.\footnote{The surface tension may be estimated as
$$
    \sigma=\int_{-\infty}^{\infty}\d r\, \left[\frac{1}{2}\left(\frac{\d \bar{\varphi}_{\rm bubble}}{\d r}\right)^2+ V^{T=0}(\bar{\varphi}_{\rm bubble})\right]\,,
$$
where $\bar{\varphi}_{\rm bubble}$ is the bubble field configuration in the rest frame of the wall.}

The fraction of the vacuum energy transferred to the wall kinetic energy is then given by
\begin{align}
    \kappa_{\rm wall}=\frac{E_{\rm wall}}{E_{\rm vac}}=\frac{3 \sigma \gamma_w \beta}{\Delta V (8\pi)^{\frac{1}{3}} v_w}\approx \frac{3 \sigma \beta}{\Delta V (8\pi)^{\frac{1}{3}}}\times \gamma_w \,,
\end{align}
where in the last step we have used $v_w\sim 1$.
To formulate things in terms of the usual phase transition parameters, we introduce the phase transition strength
\begin{align}
\label{eq:alpha-n}
    \alpha_n=\frac{\Delta V}{\rho_{\rm rad}}=\frac{30\Delta V}{ g_{\star}(T_n) \pi^2 T_n^4}\,,
\end{align}
where $\rho_{\rm rad}$ is the energy density of radiation in the universe, which can be written in terms of $g_\star(T)$, the number of relativistic degrees of freedom, and $T_{n}$ is the nucleation temperature of the phase transition. Assuming that the phase transition occurs in a radiation-dominated Universe, the Hubble parameter, $H$, is given by
\begin{align}
\label{eq:Friedmann}
    H^2=\frac{8\pi^3 g_\star(T_n)\, T_n^4}{90 M^2_{\rm Pl}}\,,
\end{align}
where the Planck mass $M_{\rm Pl}\approx 1.22\times 10^{19}$ ${\rm GeV}$. Finally, we can write $\kappa_{\rm wall}$ as
\begin{align}
    \kappa_{\rm wall}\approx 50\times \gamma_w\times\left(\frac{T_n}{M_{\rm Pl}}\right) \left(\frac{\sigma}{T_n^3}\right) \left(\frac{\beta/H}{100}\right)\left(\frac{1}{\alpha_n}\right)\left(\frac{10}{\sqrt{g_\star(T_n)}}\right)\,.
\end{align}
The condition $\kappa_{\rm wall}<1$ gives the upper bound, $\bar{\gamma}_w$,
\begin{align}
   \gamma_w<\bar{\gamma}_w\approx 2.4\times 10^{15}\times\left(\frac{100 {\rm GeV}}{T_n}\right)\left(\frac{T^3_n}{\sigma}\right)\left(\frac{100}{\beta/H}\right)\left(\frac{\alpha_n}{1}\right)\left(\frac{\sqrt{g_\star(T_n)}}{10}\right)\,.
\end{align}
$\bar{\gamma}_w$ can be very large as long as the phase transition scale is far below the Planck scale, opening a window of possibility for relativistic wall velocities.

Since $\kappa_{\rm wall}+\kappa_{\rm plasma}=1$, we use the rough criterion 
\begin{align}
    \kappa_{\rm wall} <\frac{1}{2} 
\end{align}
for when the dark matter production from bubble expansion dominates over that from bubble collision. According to this criterion, the bubble expansion production mechanism is likely more important when $\gamma_w\leq \bar{\gamma}_w/2$, a situation which can be easily realized considering the large order of magnitude of the Lorentz factors achievable.

\section{Scalar dark matter production from bubble expansion}
\label{sec:scalar}

We first start with the well-studied case of scalar production in FOPTs. This will allow us to establish our notation and methodology for the numerical calculation and for the fit that we validate against previously known results, before moving on to vector pair production in the following Section. 

\subsection{Model and particle production mechanism}
To be specific, we consider the following model
\begin{align}
\label{eq:model}
\L=
	\frac{1}{2} (\partial_\mu \Phi) (\partial^\mu \Phi)
	+ \frac{1}{2} (\partial_\mu \chi) (\partial^\mu \chi)
	+\frac{\mu^2}{2} \Phi^2
	- \frac{m_\chi^2}{2} \chi^2
	- \frac{\lambda_\phi}{4!} \Phi^4
	- \frac{\lambda_\chi}{4!} \chi^4
	- \frac{\lambda}{4} \Phi^2 \chi^2+\Delta \L\,,
\end{align}
where $\Delta\L$ describes interactions between $\Phi$ and the SM fields. The $Z_2$ symmetry ensures that $\chi$ can be a stable dark matter candidate~\cite{Silveira:1985rk,McDonald:1993ex,Burgess:2000yq,Fairbairn:2013uta,Lebedev:2021xey}. 
We assume that the $\Phi$ field undergoes a FOPT so that its vacuum expectation value is non-vanishing, $\langle \Phi(x)\rangle =v(x)$ where $v(x)$ describes the bubble background.\footnote{Note that the $\chi$ field may also form a condensate in the early Universe, although we do not consider this possibility in this work. For the evolution of such a $Z_2$ scalar background field in the early Universe, see, e.g., Refs.~\cite{Mukaida:2013xxa,Ai:2021gtg,Wang:2022mvv,Ai:2023ahr,Ai:2023qnr}.} 
In our analysis, we consider a planar wall expanding along the negative $z$-direction. In the wall's rest frame, the function $v(x)$ typically adopts the following form:
\begin{align}
\label{eq:wall-profile}
    v(z)=\frac{v_b}{2}\left[{\rm tanh}(z/L_w)+1\right]\,,
\end{align}
where $v_b>0$ is the symmetry broken value of $\Phi(x)$, and $L_w$ characterises the wall width. 

\begin{figure}[t!]
    \centering
    \includegraphics[scale=0.2]{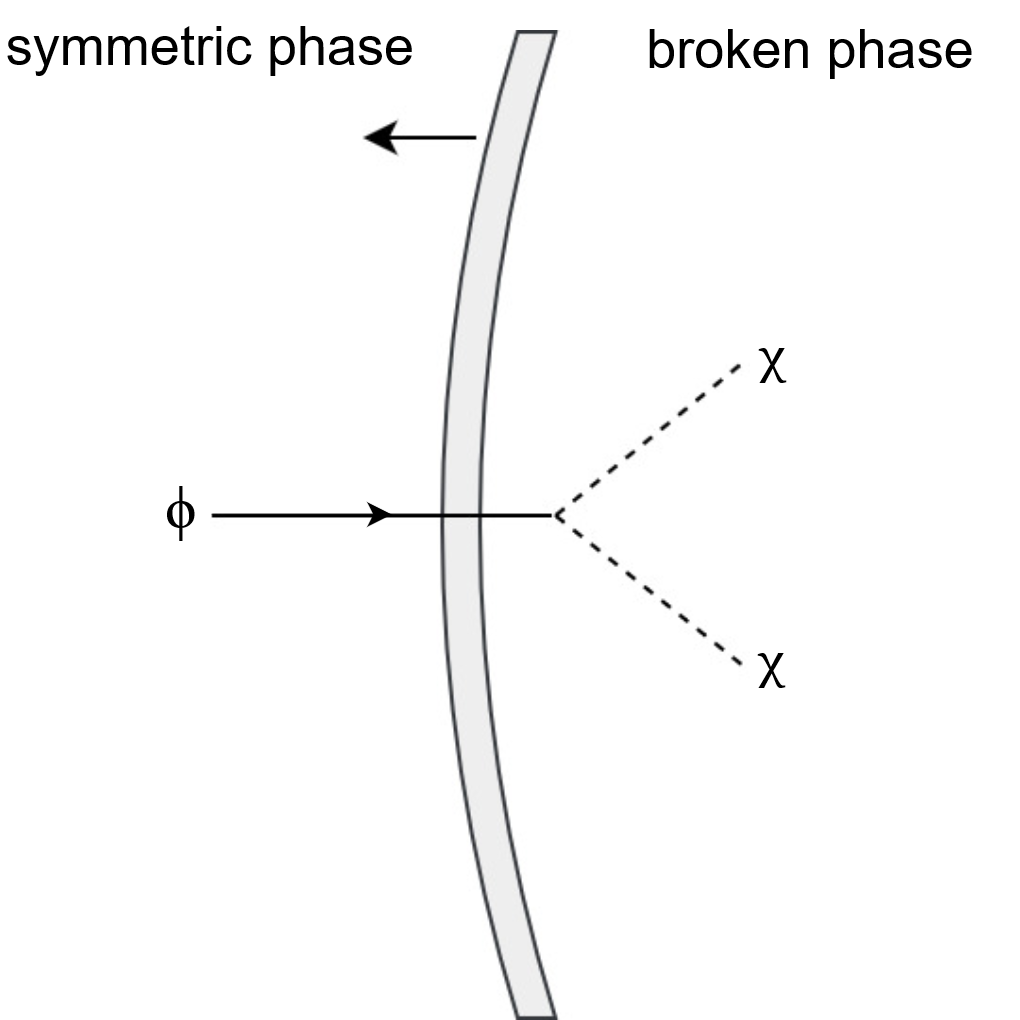}
    \caption{Schematic illustration of the light-to-heavy $1\rightarrow 2$ process in a FOPT. The grey surface represents the bubble wall of the expanding broken phase (right side of the wall) where $\langle\Phi\rangle\neq0$. A $\phi$ excitation in the plasma passing through the fast bubble wall decays into a pair of much heavier $\chi$ particles, taking $z$-momentum from the wall itself.}
    \label{fig:light-to-heavy}
\end{figure}

Substituting $\Phi(x)=v(z)+\phi(x)$ into Eq.~\eqref{eq:model}, one obtains a Lagrangian for the fluctuation fields $\phi$ and $\chi$ with masses and interactions depending on the background $v(z)$. For example, the zero-temperature mass of the $\phi$-field is $m_\phi^2(z)=-\mu^2+\lambda_\phi v^2(z)/2$. The term $\lambda \Phi^2\chi^2/4$ induces an interaction 
\begin{align}
    \L\supset -\frac{\lambda}{2} v(z) \phi \chi^2\,,
\end{align}
which gives rise to the transition $\phi\rightarrow 2\chi$. We consider the production of heavy dark matter particles from relativistic bubble walls so that we have
\begin{align}
   m_\chi \gg  (T\sim v_b\sim m_\phi) \,,
\end{align}
where $T$ is the temperature at the phase transition. With this hierarchy, the transition $\phi\rightarrow 2\chi$ is forbidden in the absence of the wall due to the inability to fulfil energy-momentum conservation. However, the bubble wall spontaneously breaks translational symmetry along the $z$-direction, and therefore the total $z$-momentum of the particles is not required to be conserved, rendering the transition $\phi\rightarrow 2\chi$ feasible. This process is schematically described in Fig.~\ref{fig:light-to-heavy}.

Because of the aforementioned hierarchy, we can ignore the field-dependent and thermal masses for the $\chi$-particles. 
To make the light-to-heavy transition possible, the $\phi$-particle must have a very large $z$-momentum. For these large $z$-momentum $\phi$-particles, we can also ignore the field-dependent and thermal masses for $\phi$ and treat them as massless. We write the four-momenta for the process under study as
\begin{subequations}
\label{eq:kinematics}
\begin{align}
    &\phi:\quad\ \  p=(E^{(\phi)}_\vecp,\vecp_\perp,p^z)\,,\\
    &\chi_1:\quad k_1=(E^{(\chi)}_{\veck_1},\veck_{1,\perp},k_1^z)\,,\\
    &\chi_2:\quad k_2=(E^{(\chi)}_{\veck_2},\veck_{2,\perp},k_2^z)\,,
\end{align}
\end{subequations}
where $\vecp_\perp= (\vecp^x,\vecp^y)$ and similarly for $\veck_{1,\perp}$, $\veck_{2,\perp}$, and $E_\vecp^{(\phi)}=|\vecp|$, $E_\veck^{(\chi)}=\sqrt{m_\chi^2+\veck^2}$.

In the wall's rest frame, the flux impinging on the wall is
\begin{align}
    J^{\rm (wall)}= \int\frac{\d^3 \vecp}{(2\pi)^3} \frac{p^z}{E_\vecp^{(\phi)}} f_\phi(\vecp,T)\,,
\end{align}
where the distribution function of $\phi$ particles is
\begin{align}
    f_\phi(\vecp,T)=\frac{1}{\e^{\frac{\gamma_w\left(E^{(\phi)}_\vecp-v_w p^z\right)}{T}}-1}\,.
\end{align}

To derive the density of the produced $\chi$ particles, we consider a duration $\Delta t$ in the wall frame. The total number of particles produced in this time is 
\begin{align}
\label{eq:particle-number}
    N = {\rm wall\ area}\times \Delta t \times 2\times\int\frac{\d^3 \vecp}{(2\pi)^3} \frac{p^z}{E_\vecp^{(\phi)}} \d\mathbb{P}_{\phi\rightarrow 2\chi}(\vecp)\times f_\phi(\vecp,T)\,.
\end{align}
where $\d\mathbb{P}_{\phi\rightarrow 2\chi}$ is the differential probability for $\phi\rightarrow 2\chi$ and the factor of $2$ is due to two $\chi$-particles being produced in one process.  In the plasma frame, the volume swept by the wall is 
\begin{align}
   \Delta V^{\rm (plasma)} = {\rm wall\ area}\times v_w\times \gamma_w\Delta t\,.
\end{align}
Dividing $N$ by $\Delta V^{\rm (plasma)}$ then gives the density of produced $\chi$-particles in the {\it plasma frame}, 
\begin{align}
\label{eq:production-rate}
    n_\chi= \frac{2}{v_w\gamma_w}\int\frac{\d^3 \vecp}{(2\pi)^3} \frac{p^z}{E_\vecp^{(\phi)}}\d\mathbb{P}_{\phi\rightarrow 2\chi}(\vecp)\times f_\phi(\vecp,T)\,.
\end{align}

The differential probability is given by~\cite{Ai:2023suz}
\begin{align}
\label{eq:dP}
    \d\mathbb{P}_{\phi\rightarrow  2\chi}(\vec{p})=\frac{\lambda^2 v_b^2 L_{w}^2  \pi^2 }{ 16  p^z}  &\prod_{i=1,2}\int\frac{\d^3\veck_i}{(2\pi)^3 2 E^{(\chi)}_{\veck_i}} \times (2\pi)^3 \delta(E^{(\phi)}_{\vecp}-E^{(\chi)}_{\veck_1}- E^{(\chi)}_{\veck_2})\delta^{(2)}(\vecp_\perp-\veck_{1,\perp}-\veck_{2,\perp})\notag\\
    &\times {\rm csch}^2\left(\frac{\pi L_w \Delta p^z }{2}\right) \,,
\end{align}
where $\Delta p^z=p^z-k_1^z-k_2^z$. The hyperbolic cosecant function originates from the Fourier transform
of the wall profile. Above, we have corrected the expression in Ref.\,\cite{Ai:2023suz} with an additional factor of $2\pi$, which was caused by a missing factor of $\sqrt{2\pi}$ in the Fourier transform $\widetilde{v}(q^z)$ used in Ref.\,\cite{Ai:2023suz}. Integrating over $\veck_1$ in the above equation and relabeling $\veck_2$ as $\veck$, we obtain
\begin{align}
    \d \mathbb{P}_{\phi\rightarrow 2\chi}(\vecp)=\frac{\lambda^2 v_b^2 L_w^2 \pi^2 }{ 32 p^z}\int\frac{\d^3 \veck}{(2\pi)^3 2E_\veck^{(\chi)}} \frac{1}{\sqrt{H(\vecp,\veck)}}\times {\rm csch}^2 \left(\frac{\pi L_w  \overline{\Delta p^z}}{2}\right)\,,
\end{align}
where 
\begin{align}
    \overline{\Delta p^z}\equiv p^z-k^z-\sqrt{H(\vecp,\veck)}\,,
\end{align}
is the on-shell $z$-momentum transfer, and 
\begin{align}
    H(\vecp,\veck)\equiv (p^z)^2+(k^z)^2+2\vecp_\perp\cdot \veck_{\perp}-2|\vecp|E_{\veck}^{(\phi)}\,.
\end{align}
In the above equations, there is the constraint $H(\vecp,\veck)\geq 0$ which is enforced by kinematics.

Following Ref.~\cite{Ai:2023suz}, one can make the following two further simplifications:
\begin{itemize}
    \item[(i)] Integrate over $p^z$ using the method of steepest descent, making use of the fact that the distribution function $f_{\phi}(\vecp,T)$ is exponentially suppressed away from the stationary point $p^z=\gamma_w|\vecp_\perp|$.
    \item[(ii)] Set a cutoff $\pi L_w\overline{\Delta p^z}/2\leq a$, making use of the fact the hyperbolic cosecant function is exponentially suppressed for large $\pi L_w\overline{\Delta p^z}/2$. Taking $a=10$ already gives a very high accuracy.
\end{itemize}
We refer the reader to Ref.~\cite{Ai:2023suz} for more details of this simplification procedure. {For $\d\mathbb{P}_{\phi\rightarrow 2\chi}(\vecp)$, we obtain
\begin{align}
\label{eq:dP-integrated-over-k1}
    \d \mathbb{P}_{\phi\rightarrow 2\chi}(|\vecp_\perp|,\gamma_w|\vecp_\perp|)= &\frac{\lambda^2 v_b^2 L_w^2}{ 256 \gamma_w |\vecp_\perp|}\int_0^{|\veck_\perp|_{\rm max}(|\vecp_\perp|) }\d |\veck_\perp|\, |\veck_\perp|\notag\\
    &\times \int_{-|k^z|_{\rm max}(|\vecp_\perp|,|\veck_\perp|)}^{|k^z|_{\rm max}(|\vecp_\perp|,|\veck_\perp|)}\d k^z\, \frac{1}{E_\veck^{(\chi)}}\frac{1}{\sqrt{G(|\vecp_\perp|,|\veck_\perp|,k^z)}} {\rm csch}^2 \left(\frac{\pi L_w \overline{\Delta p^z}}{2}\right)\,,
\end{align}}
where
\begin{subequations}
\begin{align}
\label{eq:new-integral-limits}
|\vecp_\perp|_{\rm min} &= \frac{\pi L_w m_\chi^2}{a\gamma_w}\,,\notag\\
|\veck_\perp|_{\rm max}(|\vecp_\perp|) &= \sqrt{\frac{ a\gamma_w |\vecp_\perp|}{\pi L_w} - m_\chi^2 } \,,\notag\\
|k^z|_{\rm max}(|\vecp_\perp|,|\veck_\perp|) &= \sqrt{\gamma^2_w \vecp_\perp^2 -2\gamma_w |\vecp_\perp| \sqrt{\veck_\perp^2+m^2_\chi} }\,,
\end{align}  
\end{subequations}
and
\begin{align}
 G(|\vecp_\perp|,|\veck_\perp|,k^z) \equiv \gamma_w^2 \vecp_\perp^2+(k^z)^2-2\gamma_w |\vecp_\perp| E^{(\chi)}_\veck\,.
\end{align}
Now the on-shell $z$-momentum transfer reads
\begin{align}
    \overline{\Delta p^z}= \left(\gamma_w |\vecp_\perp|-k^z-\sqrt{G(|\vecp_\perp|,|\veck_\perp|,k^z)}\right)\,.
\end{align}
The final result for $n_\chi$ is
\begin{align}
\label{eq:master-integral}
    n_\chi=\frac{\sqrt{2\pi T}\lambda^2 v_b^2 L_w^2}{ 512\pi^2 }&\frac{1}{\gamma_w} \int_{|\vecp_\perp|_{\rm min}}^\infty \d |\vecp_\perp|\, |\vecp_\perp|^{1/2} \e^{-\frac{|\vecp_\perp|}{T}}\int_0^{|\veck_\perp|_{\rm max}(|\vecp_\perp|) }\d |\veck_\perp|\, |\veck_\perp|\notag\\
    &\times\int_{-|k^z|_{\rm max}(|\vecp_\perp|,|\veck_\perp|)}^{|k^z|_{\rm max}(|\vecp_\perp|,|\veck_\perp|)}\d k^z\, \frac{1}{E_\veck^{(\chi)}}\frac{1}{\sqrt{G(|\vecp_\perp|,|\veck_\perp|,k^z)}} {\rm csch}^2 \left(\frac{\pi L_w \overline{\Delta p^z}}{2}\right)\,,
\end{align}

\subsection{Numerical results and fit}
\label{sec:scalar-fit}

In this subsection, we compute the integral~\eqref{eq:master-integral} numerically. Defining the following dimensionless variables
\begin{align}
\label{eq:dimensionalless-variables}
    x=\frac{|\vecp_\perp|}{T}\,,\qquad y=\frac{|\veck_\perp|}{T}\,,\qquad z=\frac{k^z}{T}\,,\qquad \xi=\frac{m_\chi}{T}\,,\qquad \kappa =L_w T\,,
\end{align}
{the integral~\eqref{eq:dP-integrated-over-k1} can be written as
\begin{align}
\label{eq:dP-simplified}
    \d \mathbb{P}_{\phi\rightarrow 2\chi}(|\vecp_\perp|,\gamma_w|\vecp_\perp|)= &\frac{\lambda^2 (v_b/T)^2 \kappa^2 }{ 256\gamma_w  x}\int_0^{y_{\rm max}(x) }\d y\, y\int_{-|z|_{\rm max}(x,y)}^{|z|_{\rm max}(x,y)}\d z\, \frac{{\rm csch}^2 \left(\frac{\kappa\pi \left[\gamma_w x-z-\widetilde{G}^{1/2}\right]}{2}\right)}{\widetilde{E}\widetilde{G}^{1/2}}\,,
\end{align}
where 
\begin{align}
    \widetilde{E}=\sqrt{\xi^2+y^2+z^2}\,,\qquad \widetilde{G}=\gamma_w^2 x^2 +z^2 -2\gamma_w x \widetilde{E}\,,
\end{align}
and
\begin{align}
    y_{\rm max}(x)=\sqrt{\frac{a \gamma_w x}{\pi\kappa}-\xi^2}\,,\qquad |z|_{\rm max}(x,y)=\sqrt{\gamma_w^2 x^2-2\gamma_w x \sqrt{\xi^2+y^2} }\,.
\end{align}
As expected, $\d\mathbb{P}_{\phi\rightarrow 2\chi}$ is dimensionless.}
The integral~\eqref{eq:master-integral} can be written as
\begin{align}
    \label{eq:master-integral2}
    n_\chi&=\frac{\sqrt{2\pi} \lambda^2 v_b^2 \kappa^2 T^3}{512\pi^2 m_\chi^2}\times I(\xi,\gamma_w,\kappa)\,, \\
    I(\xi,\gamma_w,\kappa)&\equiv \frac{\xi^2}{\gamma_w}\int_{x_{\rm min}}^\infty \d x\, x^{1/2} \e^{-x}\int_0^{y_{\rm max}(x)}\d y\, y \int_{-|z|_{\rm max}(x,y)}^{|z|_{\rm max}(x,y)}\d z\, \frac{{\rm csch}^2 \left(\frac{\kappa\pi \left[\gamma_w x-z-\widetilde{G}^{1/2}\right]}{2}\right)}{\widetilde{E}\widetilde{G}^{1/2}}\notag \, , 
\end{align}
{where $x_{\rm min}=\frac{\pi\kappa\xi^2}{a\gamma_w}$.}
Note that we have absorbed the explicit factor of $\xi^2/\gamma_w$ in Eq.~\eqref{eq:master-integral2} into the definition of the integral $I$. For the convenience of future use, we denote
\begin{align}
\label{eq:factorA}
    A\equiv\frac{\sqrt{2\pi} \lambda^2 v_b^2 \kappa^2 T^3}{ 512\pi^2  m_\chi^2 }\,.
\end{align}
We shall also compare our results with those given in Ref.~\cite{Azatov:2021ifm}. There, $n_\chi$ is estimated as\footnote{Note that the $\lambda$ in this work is double that of Ref.~\cite{Azatov:2021ifm} and that a factor of 2 in the expression of Ref.~\cite{Azatov:2021ifm} has been corrected in the recent paper~\cite{Azatov:2024crd}.}
\begin{align}
\label{eq:AVY}
    n_{\chi;\rm AVY}=\frac{T^3\lambda^2 v_b^2}{96\pi^4 m^2_\chi}\, \e^{-\frac{1}{2}\frac{L_w T}{\gamma_w}\frac{m^2_\chi}{T^2}}=A \times \left(\frac{32 }{3 \pi}  \frac{1}{\sqrt{2\pi^3}} \frac{1}{\kappa^2} \,\e^{-\frac{1}{2}\frac{L_w T}{\gamma_w}\frac{m^2_\chi}{T^2}}\right)\,.
\end{align}
Next, we perform our numerical calculation and fit for a fixed value of $\kappa \equiv L_w T$ before fitting the $\kappa$-dependence in $I(\xi,\gamma_w,\kappa)$.  

\begin{figure}[H]
    \centering
    \includegraphics[scale=0.6]{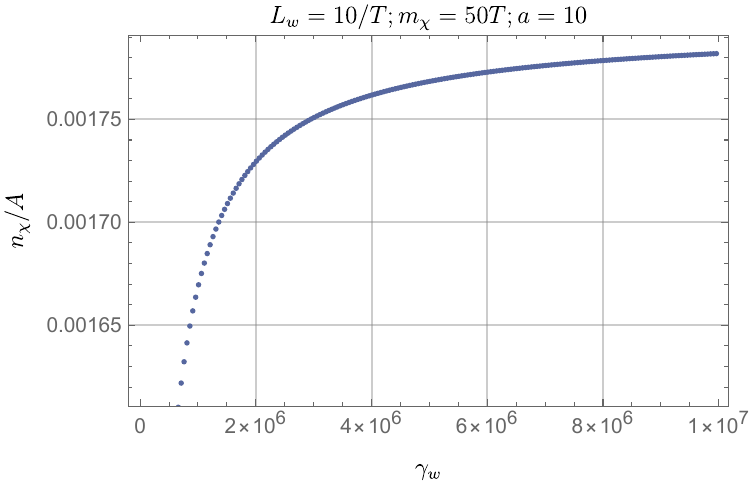}
    \includegraphics[scale=0.6]{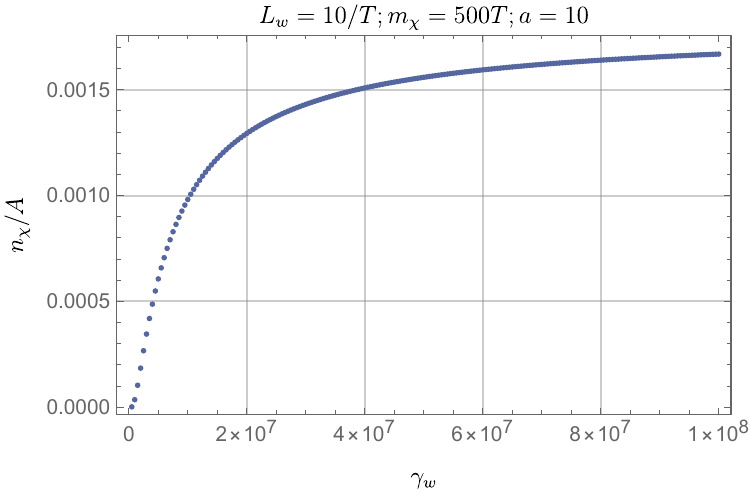}
    \includegraphics[scale=0.6]{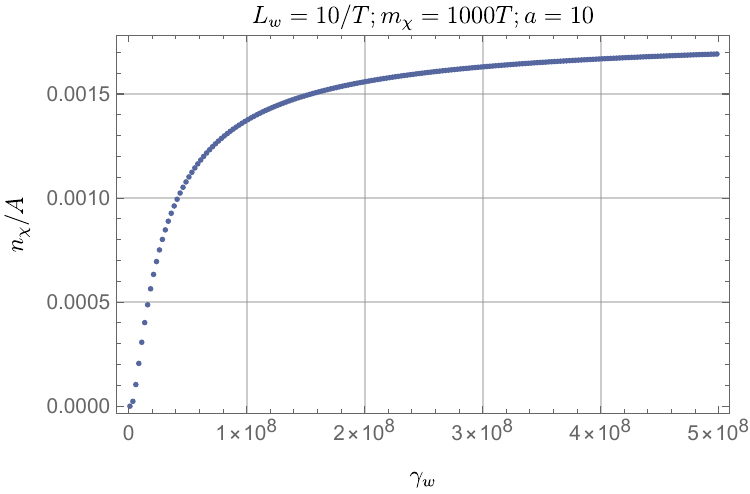}
    \caption{Numerical results of $n_\chi/A$ as a function of $\gamma_w$ for fixed $m_\chi/T=50, 500, 1000$, respectively.}
    \label{fig:nphi}
\end{figure}

\paragraph{Results for fixed $\bm{\kappa=10}$}--- A typical value of $\kappa$ at
the electroweak phase transition with or without supersymmetry is of $\O(10)$~\cite{Moore:1995si,Moreno:1998bq}. In this subsection we fix $\kappa=10$. Let us first look at the behaviour of the integral $I(\xi,\gamma_w,\kappa=10)$ as a function of $\gamma_w$ for fixed values of $m_\chi/T$. In Fig.~\ref{fig:nphi}, we show $n_\chi/A$ as a function of $\gamma_w$ for fixed $m_\chi=50T, 500 T, 1000 T$, respectively. 

\begin{figure}[H]
    \centering
    \includegraphics[scale=0.6]{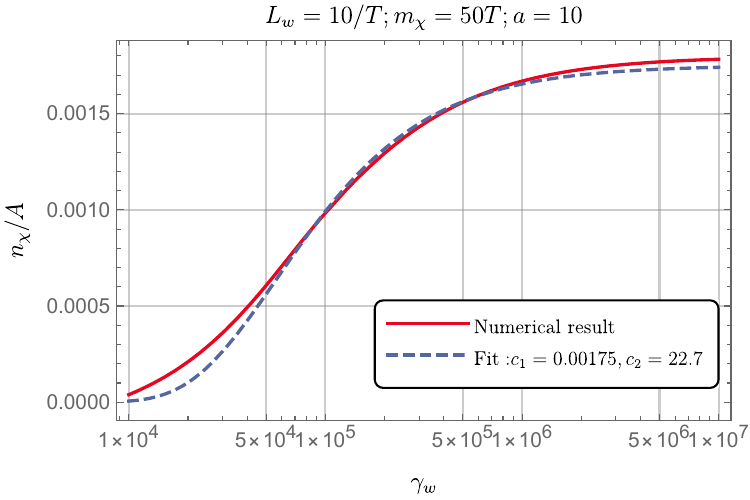}
    \includegraphics[scale=0.6]{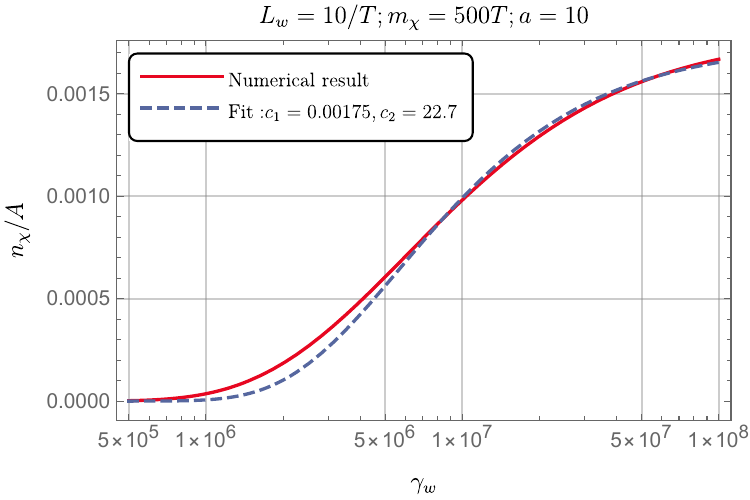}
     \includegraphics[scale=0.6]{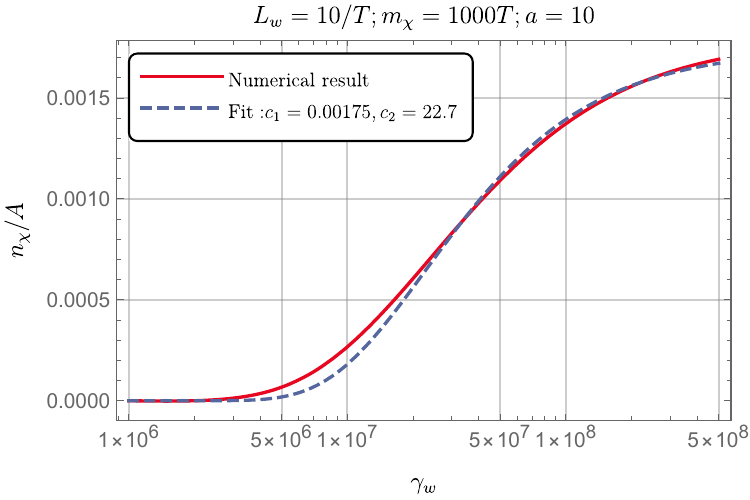}
    \caption{Comparison between the numerical results presented in Fig.~\ref{fig:nphi} with the fit function given in Eq.~\eqref{eq:fit-function} with $c_1=0.00175, c_2=22.7$.}
    \label{fig:nphi-fit}
\end{figure}

We fit the numerical results with the following expression
\begin{align}
    \label{eq:fit-function}
    I(\xi,\gamma_w,\kappa=10)=c_1 \, \e^{-\frac{c_2}{\gamma_w} \frac{m_\chi^2}{T^2} }\,.
\end{align}
We find the best fit $c_1=0.00175$, $c_2=22.7$. The numerical results and the fit are compared in Fig.~\ref{fig:nphi-fit}. The AVY formula Eq.~(\ref{eq:AVY}) corresponds to $c_1=0.02709$, $c_2=5$.

\begin{figure}[H]
    \centering
     \includegraphics[scale=0.6]{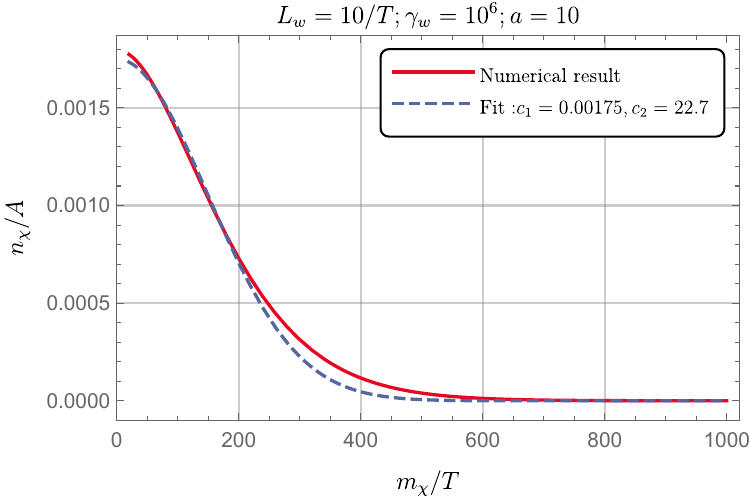}
    \caption{Numerical result of $n_\chi/A$ in the range $m_\chi/T\in [20,1000]$ for $\gamma_w=10^6$, as well as its comparison between with the fit function given in Eq.~\eqref{eq:fit-function} with $c_1=0.00175, c_2=22.7$.}
    \label{fig:nphi4}
\end{figure}

We can use the numerical result of $n_\chi/A$ for a fixed $\gamma_w$ as a cross-check of the fit.  In Fig.~\ref{fig:nphi4}, we show  $n_\chi/A$ as a function of $m_\chi/T$ as well as its comparison with the fit. We see that the fit function slightly underestimates the number density in the range $m_\chi/T$ between 200 and 500 but provides a good overall parameterisation. 

\begin{figure}[H]
    \centering
    \includegraphics[scale=0.6]{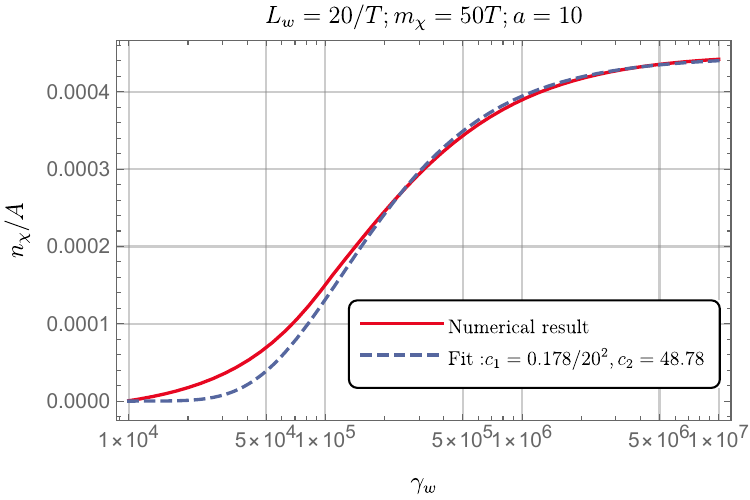}
    \includegraphics[scale=0.6]{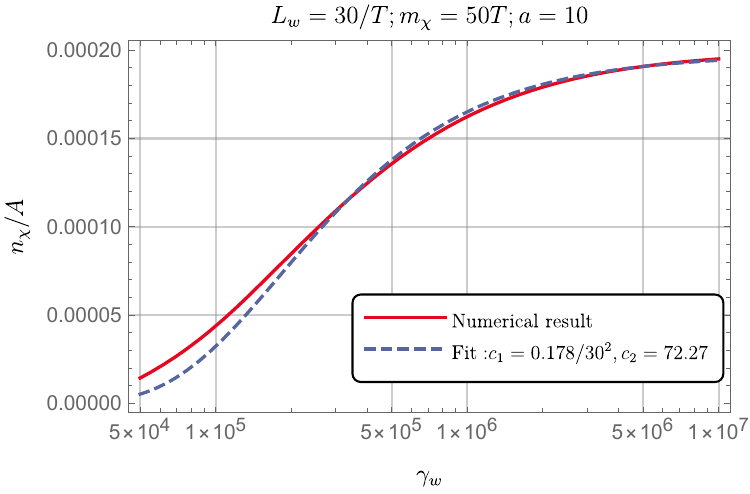}
    \includegraphics[scale=0.6]{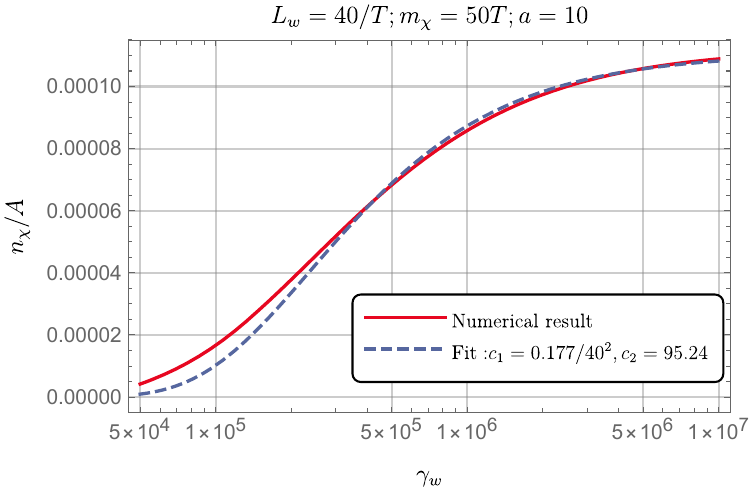}
    \includegraphics[scale=0.6]{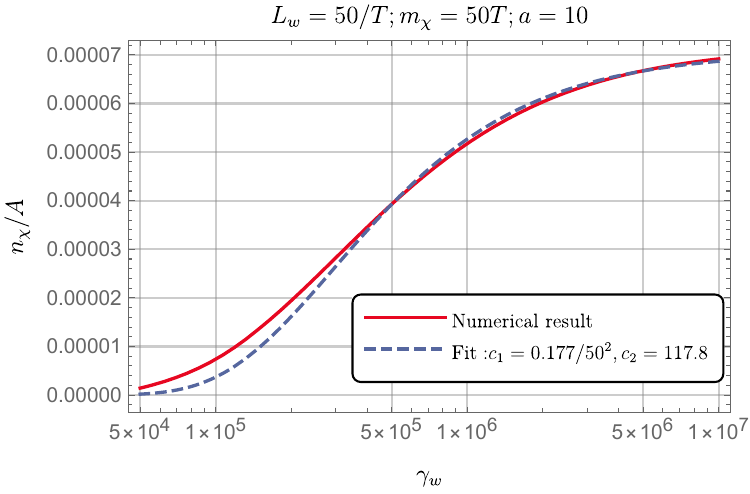}
    \caption{Comparison between the numerical results for different values of $L_w T$ and the fit using parameters given in Table~\ref{tab:fit-kappa}.}
    \label{fig:fit-kappa}
\end{figure}

\paragraph{Results fitting the dependence on $\kappa$} --- Now we give the fit for the dependence of $I(\xi,\gamma_w,\kappa)$ on $\kappa$. For simplicity, we fix $m_\chi=50 T$. The best fit parameters $c_1$ and $c_2$ for $\kappa=10,...,50$ are given in Table~\ref{tab:fit-kappa}. We show the comparison between the numerical results and fit function in Fig.~\ref{fig:fit-kappa}. From the table, it is easy to deduce that 
\begin{align}
    c_1 = 0.177\times \frac{1}{\kappa^2}\,,\quad c_2=2.37\times \kappa\,. 
\end{align}

\begin{table*}[ht]
\centering
    \begin{tabular}{|c|c|c|c|c|c|}
    \hline
    & $\kappa=10$   & $\kappa=20$ & $\kappa=30$ & $\kappa=40$ & 50 \\
    \hline         
   $c_1$ & $0.175/10^2$ & $0.178/20^2$ & $0.178/30^2$ & $0.177/40^2$ & $0.177/50^2$\\
    \hline
    $c_2$ & $22.7$ & $48.78$ &  $72.27$ & $95.24$ & $117.8$\\
    \hline
    \end{tabular}
    \caption{Fitted parameters of $c_1$ and $c_2$ for different values of $\kappa$.}
    \label{tab:fit-kappa}
\end{table*}

Finally, the fit function for the produced dark matter density in the plasma frame reads
\begin{align}
\label{eq:nchi}
    \boxed{n_\chi\approx 0.88 \times 10^{-4} \times \frac{\lambda^2 v_b^2  T^3}{m_\chi^2}\times  \e^{-2.37 \frac{L_w T}{\gamma_w} \frac{m_\chi^2}{T^2}}\,.}
\end{align}
For comparison, the AVY estimate reads~\cite{Azatov:2021ifm} 
\begin{align}
\label{eq:nchi-AVY}
    n_{\chi;\rm AVY}\approx 1.07\times 10^{-4}\times\frac{\lambda^2 v_b^2 T^3 }{m^2_\chi}\, \e^{-\frac{1}{2}\frac{L_w T}{\gamma_w}\frac{m^2_\chi}{T^2}}\,.
\end{align}
{We note that the analytic result~\eqref{eq:nchi-AVY} is derived under a few mild assumptions. For example, instead of doing the integral using the tangent wall profile throughout, the latter is replaced by its limit for $L_w\overline{\Delta p^z}\rightarrow 0$ multiplied with a hard cutoff given by a Heaviside step function.  Another issue is that Ref.~\cite{Azatov:2021ifm} assumes $\vecp_\perp=0$ when calculating $\d \mathbb{P}_{\phi\rightarrow 2\chi}(\vecp)$. This itself is not a problem because one can always make $\vecp_\perp=0$ with a transverse boost. However, when substituting $\d \mathbb{P}_{\phi\rightarrow 2\chi}(\vecp)$ into $n_\chi$, one needs to take an inverse transverse boost to recover the $|\vecp_\perp|$-dependence in $\d \mathbb{P}_{\phi\rightarrow 2\chi}(\vecp)$ and then integrate over $|\vecp_\perp|$ in $n_\chi$. The second step is not done in Ref.~\cite{Azatov:2021ifm}. It is therefore not surprising to see a minor discrepancy between our numerical result~\eqref{eq:nchi} and the analytic result given in Ref.~\cite{Azatov:2021ifm}.}

Assuming no additional processes that decrease the dark matter density except for redshift, one obtains the relic abundance of the scalar dark matter today,
\begin{align}
    \overbar{\Omega}_{\chi,\rm BE}^{\rm today} h^2=\frac{m_\chi n_\chi}{\rho_c/ h^2} \frac{g_{\star S}(T_0) T_0^3 }{g_{\star S}(T_{\rm after})  T^3_{\rm after}}\,,
\end{align}
where $\rho_c$ is the critical energy density, $g_{\star S}$ is the effective degrees of freedom for entropy,  $T_0$ is the temperature today, and ($\alpha_n$ is defined in Eq.~\eqref{eq:alpha-n})
\begin{align}
    T_{\rm after}\approx (1+\alpha_n)^{1/4} T_n
\end{align}
being the temperature after the phase transition has been completed. Substituting our fit formula into the above equation, we obtain (taking $T=T_n$)
\begin{align}
\label{eq:Omegaphi}
    \boxed{\overbar{\Omega}_{\chi,\rm BE}^{\rm today} h^2\approx  5.5  \times \left(\frac{\lambda}{0.1}\right)^2\left(\frac{100 v_b}{m_\chi}\right)\left(\frac{100}{g_{\star S} (T_{\rm reh})}\right)\left(\frac{ v_b}{100 \rm GeV}\right)\left(\frac{T_n}{T_{\rm after}}\right)^3\times \e^{-2.37\frac{L_w T_n}{\gamma_w}\frac{m_\chi^2}{T_n^2}}\,.}
\end{align}
Above, $\gamma_w$ should be understood as the Lorentz factor of the terminal wall velocity. We have used $\rho_c \approx 1.05\times 10^{-5} h^2 \,{\rm GeV}\cdot {\rm cm}^{-3}$, $T_0\approx 2.7255 K$, and $g_{\star S}(T_0)\approx 3.9$. The above results may be viewed only as an upper bound since possible wash-out processes, e.g. dark matter annihilation, may occur after the production. As we shall see in Sec.~\ref{sec:wash-out}, wash-out processes could lead to a final relic abundance that is not sensitive to the initial condition generated at the phase transition, as long as dark matter is initially over-produced at the time of the phase transition.


\section{Vector dark matter production from bubble expansion}
\label{sec:vectorproduction}

\subsection{Production rate of the vector bosons}

Now we shall consider the production of massive vector particles. To do so, we adopt the following phenomenological Lagrangian as an effective parameterisation of a more UV-complete theory,
\begin{align}
\label{eq:Lagrangian-Proca}
    \L=\frac{1}{2}\partial_\mu \Phi\partial^\mu \Phi-V(\Phi) -\frac{1}{4}F_{\mu\nu}F^{\mu\nu}+\frac{1}{2}m^2_V V_\mu V^\mu+\frac{\lambda}{4} \Phi^2 V_\mu V^\mu\,,
\end{align}
where $F_{\mu\nu}=\partial_\mu V_\nu-\partial_\nu V_\mu$. This vector-Higgs portal~\cite{Lebedev:2011iq} can originate, for example, from the Higgs mixing with a charged complex scalar. The details of the UV completion may affect the particle production if the bubble wall becomes sensitive to sufficiently high scales; we assume the UV scale to be high enough for the EFT to be valid and consider here only the dimension-4 effective interaction for simplicity. Substituting $\Phi(x)=v(z)+\phi(x)$ into the interacting term between $\Phi$ and $V_\mu$, one has 
\begin{align}
    \L\supset \frac{\lambda}{4}v^2(z)V_\mu V^\mu+\frac{\lambda}{2} v(z) \phi V_\mu V^\mu\,.
\end{align}
The first term is a field-dependent mass for $V^\mu$ which is assumed to be negligible compared to $m_V$. The second term gives rise to the process $\phi\rightarrow 2V$ in bubble expansion. Again, we assume $m_V \gg T$.

Following Ref.~\cite{Ai:2023suz}, it is straightforward to obtain
\begin{align}
    &\d\mathbb{P}_{\phi\rightarrow 2 V} (\vecp)=\frac{1}{4p^z}\prod_{i=1,2}\int\frac{\d^3\veck_i}{(2\pi)^3 2 E^{(V)}_{\veck_i}}  \sum_{r,r'} (2\pi)^3 \delta(E^{(V)}_{\veck_1}+E^{(V)}_{\veck_2}-E^{(h)}_{\vecp})\delta^{(2)}(\vecp_\perp-\veck_{1,\perp}-\veck_{2,\perp})|\M_{\phi\rightarrow 2V}^{(r,r')}|^2\,,
\end{align}
where
\begin{align}
     |\M^{(r,r')}_{\phi\rightarrow 2V}|^2=\frac{\lambda^2 v_b^2 L_w^2\pi^2}{ 4 }{\rm csch}^2\left(\frac{\pi L_w\Delta p^z}{2}\right)\left(\epsilon_\mu^{(r)}(\veck_1)\epsilon_\nu^{(r)*}(\veck_1)\right)\left(\epsilon^{\mu(r')}(\veck_2)\epsilon^{\nu(r')*}(\veck_2)\right)\,.
\end{align}
Above, $r=\pm,0$ is an index for the polarization and $\epsilon_\mu^{(r)}$ are polarization vectors. 
The polarization vectors obey the sum relation
\begin{align}
    \sum_{r=\pm,0}\epsilon_\mu^{(r)}(\veck)\epsilon_\nu^{(r)*}(\veck)=-g^{\mu\nu}+\frac{k^\mu k^\nu}{k^2}\,,
\end{align}
where $k^\mu=(E^{(V)}_\veck,\veck)$. 
Therefore, we have 
\begin{align}
     \sum_{r,r'}|\M^{(r,r')}_{\phi\rightarrow 2V}|^2=\frac{\lambda^2 v_b^2 L_w^2\pi^2}{ 4 } \left(2+\frac{(k_1\cdot k_2)^2}{m^4_V}\right) {\rm csch}^2\left(\frac{\pi L_w\Delta p^z}{2}\right)\,.
\end{align}
Due to the assumption that the additional mass of the vector boson contributed from the background field is much smaller than its bare mass, $\lambda v^2(z) /2 \ll m_V^2$, we can safely ignore the impact of $z$-dependent mass term on the polarisations, which has been carefully studied in Ref.~\cite{Azatov:2023xem}.

Compared with the scalar case in Eq.~\eqref{eq:dP}, we only need to make the replacement $m_\chi\rightarrow m_V$ and insert an additional factor 
\begin{align}
\label{eq:factor}
   \left(2+\frac{(k_1\cdot k_2)^2}{m^4_V}\right)\, 
\end{align}
in the integrals. We therefore need to track the form of $k_1\cdot k_2$ at each step in the calculation outlined for the scalar particle production. 

Doing the integral over $\veck_1$ and relabelling $\veck_2$ as $\veck$, we obtain 
\begin{align}
    k_1\cdot k_2 \rightarrow (E_\vecp^{(\phi)}-E_\veck^{(V)})E_\veck^{(V)}-k^z\sqrt{H(\vecp,\veck)}-(\vecp_\perp-\veck_\perp)\cdot\veck_\perp\,.
\end{align}
Further applying the method of steepest descent to the integral over $p^z$ as explained above, $(k_1\cdot k_2)$ becomes
\begin{align}
    k_1\cdot k_2 \rightarrow \gamma_w |\vecp_\perp| \sqrt{m_V^2+\veck^2}-m_V^2-(k^z)^2-k^z\sqrt{G(|\vecp_\perp|,|\veck_\perp|,k^z)}-|\vecp_\perp||\veck_\perp|\cos\theta\,,
\end{align}
where $\theta$ is the angle between $\vecp_\perp$ and $\veck_\perp$. Finally, we can write
\begin{align}
    (k_1\cdot k_2)^2=J_1^2+2 J_1J_2\cos\theta+J_2^2\cos^2\theta\,,
\end{align}
where
\begin{subequations}
    \begin{align}
    &J_1(|\vecp_\perp|,|\veck_\perp|,k^z)=\gamma_w |\vecp_\perp| \sqrt{m_V^2+\veck^2}-m_V^2-(k^z)^2-k^z\sqrt{G(|\vecp_\perp|,|\veck_\perp|,k^z)}\,,\\
    &J_2(|\vecp_\perp|,|\veck_\perp|)=|\vecp_\perp||\veck_\perp|\,.
\end{align}
\end{subequations}
In the scalar case, the integrand does not have any $\theta$-dependence (to a good approximation) so the integral over $\theta$ simply contributes a factor of $2\pi$. In the present case, the term proportional to $\cos\theta$ vanishes in the integral over $\theta$ while the term proportional to $\cos^2\theta$ contributes to a factor of $\pi$. Therefore, we only need to insert the following factor into Eq.~\eqref{eq:master-integral},
\begin{align}
\label{eq:factor2}
    \left(2+\frac{J_1^2+J_2^2/2}{m_V^4}\right)
\end{align}
to get the integral for $n_V$. Using the dimensionless variables defined in Eq.~\eqref{eq:dimensionalless-variables} (with $m_\phi$ replaced by $m_V$), we have 
\begin{subequations}
    \begin{align}
   \frac{J_1}{m_V^2}&=\frac{1}{\xi^2} \left(\gamma_w x\sqrt{\xi^2+y^2+z^2}-\xi^2-z^2-z\widetilde{G}^{1/2}\right)\,,\\
   \frac{J_2}{m_V^2}&=\frac{xy}{\xi^2}\,.
\end{align}
\end{subequations}

\subsection{Unitarity violation of the longitudinal mode}

The Lagrangian~\eqref{eq:Lagrangian-Proca} is non-renormalizable and UV incomplete and this leads to a unitarity problem; the probability $\d\mathbb{P}_{\phi\rightarrow 2V}(\vecp)$ may increase without bound as $|\vecp|$ increases. This behaviour is caused by the emission of the longitudinal mode of the vector boson, denoted as $V_L$ below. Since the thermal distribution function for the $\phi$ particles, $f_\phi(\vecp,T)$, imposes $|\vecp_\perp|\sim T$, $p^z\sim \gamma_w T$, we would encounter potential unitarity breakdown only for very large $\gamma_w$. 

In a slightly different context and when $V_\mu$ is a gauge field, the authors of Ref.~\cite{Giudice:2024tcp} propose to use the Goldstone Equivalence Theorem (GET) to tame the unitarity problem. Specifically, Ref.~\cite{Giudice:2024tcp} proposes that 
for $\vecp^2 > m_V^2$, $\d\mathbb{P}_{\phi\rightarrow 2 V_L}(\vecp)$ is given by the differential probability for the transition of $\phi$ to two corresponding Goldstone bosons.

In our case, $V_\mu$ is not necessarily a gauge field that obtains a mass from a Higgs mechanism. Without specifying a UV complete theory, in this paper, we check that $\d\mathbb{P}_{\phi\rightarrow 2V}(|\vecp_\perp|,\gamma_w|\vecp_\perp|)\lesssim 1$ to be within the EFT regime of validity. 
Inserting the particular factor~\eqref{eq:factor2} into Eq.~\eqref{eq:dP-simplified}, we obtain
\begin{align}
    \d \mathbb{P}_{\phi\rightarrow 2V}(|\vecp_\perp|,\gamma_w|\vecp_\perp|)= &\frac{\lambda^2 (v_b/T)^2 \kappa^2 }{256\gamma_w  x}\int_0^{y_{\rm max}(x) }\d y\, y\notag\\
    &\int_{-|z|_{\rm max}(x,y)}^{|z|_{\rm max}(x,y)}\d z\, \frac{{\rm csch}^2 \left(\frac{\kappa\pi \left[\gamma_w x-z-\widetilde{G}^{1/2}\right]}{2}\right)}{\widetilde{E}\widetilde{G}^{1/2}}\times \left(2+\frac{J_1^2+J_2^2/2}{m_V^4}\right)\,.
\end{align}
For simplicity, below we take $|\vecp_\perp|=T$ (equivalently $x=1$) and $v_b=T$ when checking the condition $\d\mathbb{P}_{\phi\rightarrow 2V}\leq  1$. This way, $\d\mathbb{P}_{\phi\rightarrow 2V}/\lambda^2$ is only a function of $\gamma_w$.

\subsection{Numerical results and fit}

As before, we first consider the case of fixed $\kappa \equiv L_w T$ before presenting the fit results including the dependence on $\kappa$.

\paragraph{Results for fixed $\bm{\kappa=10}$} --- As in the case of scalar particle production, we can write
\begin{align}
    n_V=A\times I(\xi,\gamma_w,\kappa)\,,
\end{align}
where $A$ is given in Eq.~\eqref{eq:factorA} with $m_\phi$ replaced by $m_V$. For fixed $\kappa=10$, we find that $I$ can be very well fitted by the following function
\begin{align}
\label{eq:fit-function2}
    I(\xi,\gamma_w,\kappa=10)=c\times  \frac{\gamma_w T^2}{m_V^2} \quad {\rm with\quad } c=1.41\times 10^{-3}\,.
\end{align}

\begin{figure}[H]
    \centering
    \includegraphics[scale=0.6]{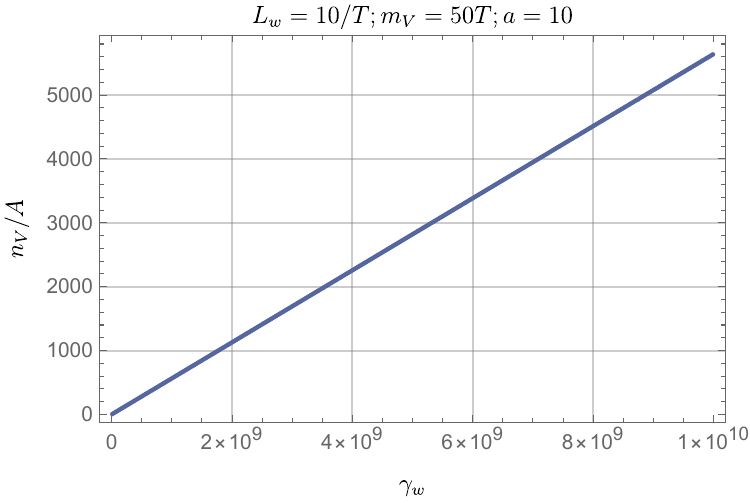}
    \includegraphics[scale=0.6]{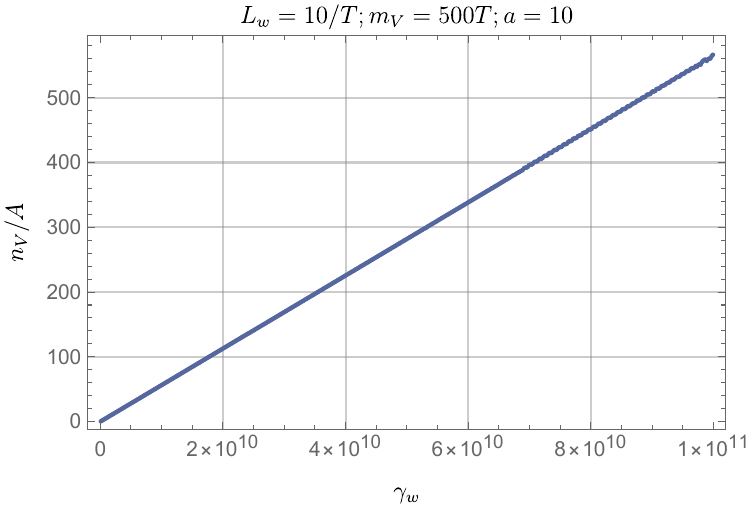}
    \includegraphics[scale=0.6]{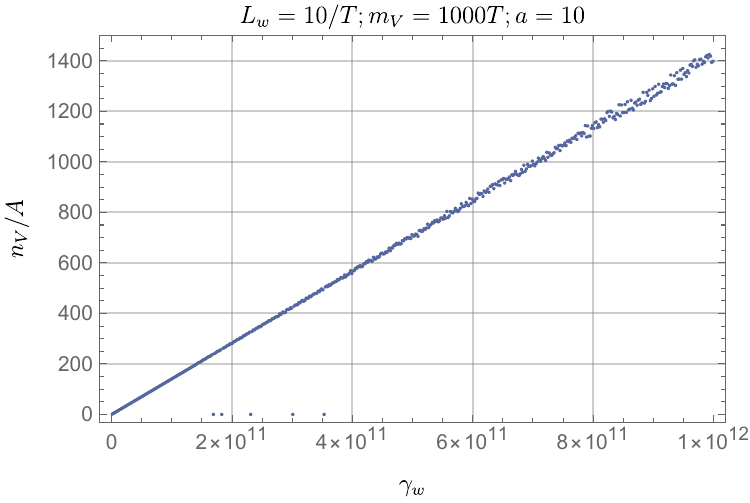}
    \caption{Numerical result of $n_V/A$ for fixed $m_V/T=50, 500, 1000$, respectively.}
    \label{fig:nV1}
\end{figure}

\begin{figure}[H]
    \centering
    \includegraphics[scale=0.65]{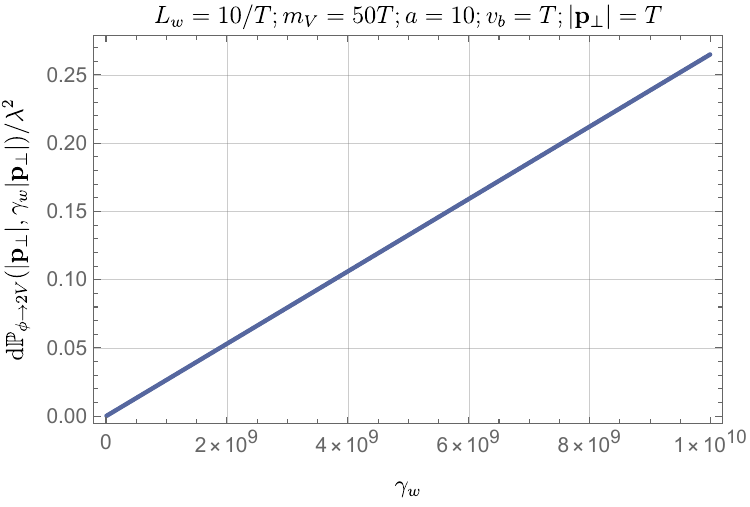}
    \includegraphics[scale=0.65]{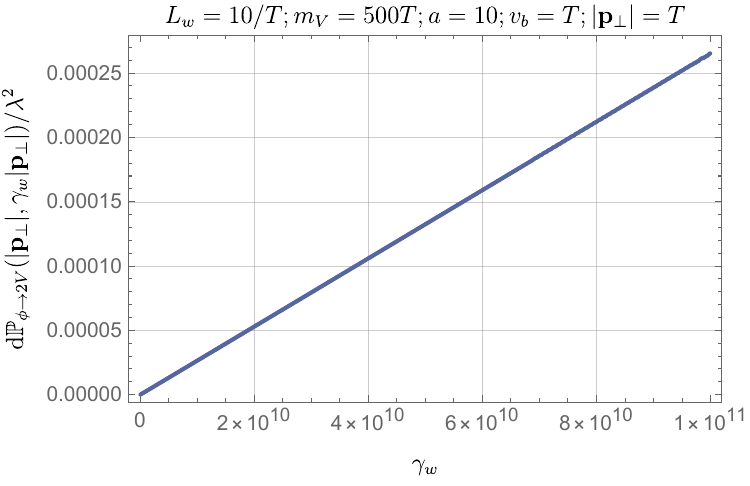}
    \caption{The probability $\d\mathbb{P}_{\phi\rightarrow 2V}(|\vecp_\perp|,\gamma_w|\vecp_\perp|)$ for fixed $m_V/T=50, 500$, respectively. We have taken $|\vecp_\perp|\sim T$ in accordance with the thermal distribution of $\phi$-particles and have also taken $v_b=T$ for simplicity.}
    \label{fig:dP1}
\end{figure}

\begin{figure}[H]
    \centering
     \includegraphics[scale=0.6]{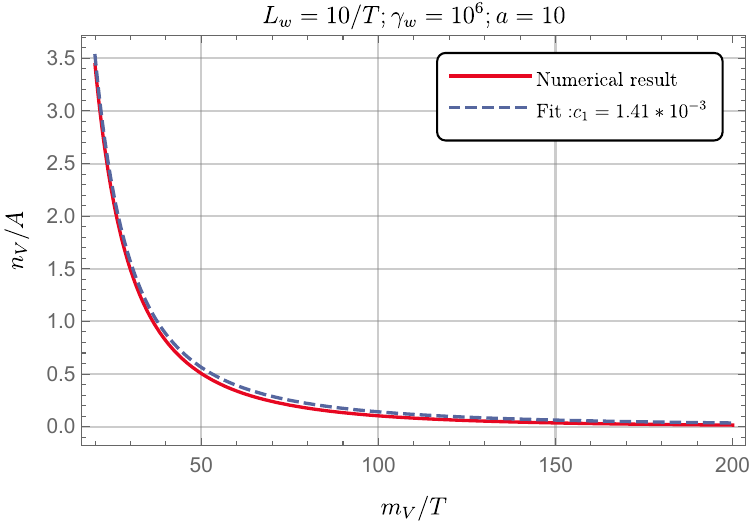}
    \caption{Numerical result of $n_V/A$ in the range $m_V/T\in [20,200]$ for $\gamma_w=10^6$ as well as its comparison with the fit function given in Eq.~\eqref{eq:fit-function2} with $c=1.41\times 10^{-3}$.}
    \label{fig:nV2}
\end{figure}

In Fig.~\ref{fig:nV1}, we show $n_V/A$ as a function of $\gamma_w$ for fixed $m_V=50T, 500 T, 1000 T$, respectively. Clearly, the numerical results show a linear behaviour in $\gamma_w$ in the regions of $\gamma_w$ studied. While this linear behaviour could originate from some other functions taken in certain limits, e.g., $\log(1+x)$ for $x\ll 1$, we checked a case with a smaller $m_V$ ($m_V=10 T$) and even higher $\gamma_w$ ($\gamma_w\in [10^{12},10^{13}]$) and confirmed that such linear behaviour persists. {We also check the unitarity condition for the scanned regions of $\gamma_w$. We find that the probability $\d\mathbb{P}_{\phi\rightarrow 2V}/\lambda^2$ indeed increases with $\gamma_w$. This means that there should be a critical value of $\gamma_w$ above which our fit formulae would no longer be valid. However, the probability is less than one for the already very large values of $\gamma_w$; see Fig.~\ref{fig:dP1}. We have checked that increasing $m_V$ or $\kappa$ would make $\d\mathbb{P}_{\phi\rightarrow 2V}/\lambda^2$ smaller. Therefore, for the phenomenological application below, our numerical results should be safe.} Finally, we compare the fit function with the numerical result as a function of $\xi$ for fixed $\gamma_w=10^6$ in Fig.~\ref{fig:nV2}.

\begin{figure}[H]
    \centering
    \includegraphics[scale=0.6]{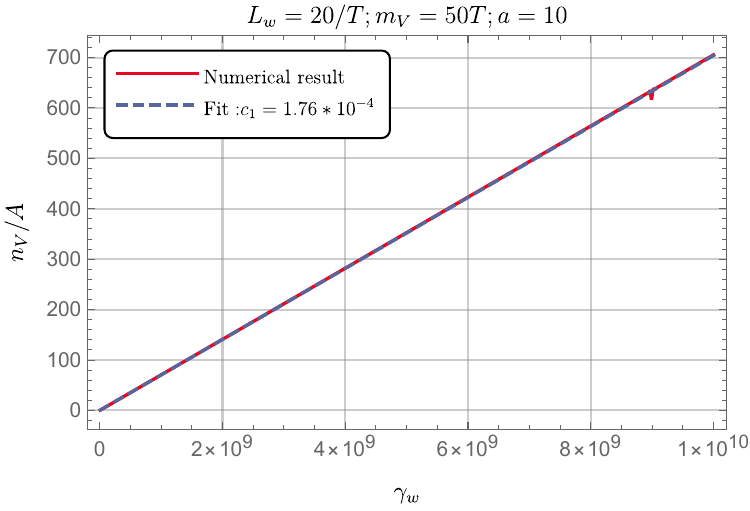}
    \includegraphics[scale=0.6]{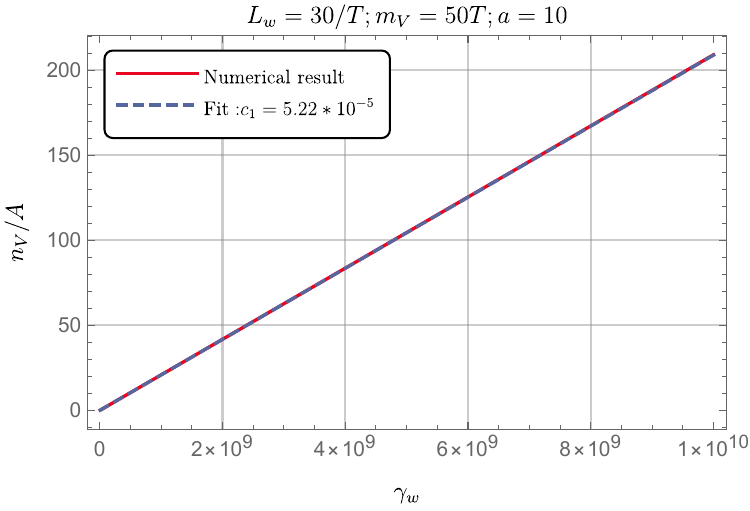}
    \includegraphics[scale=0.6]{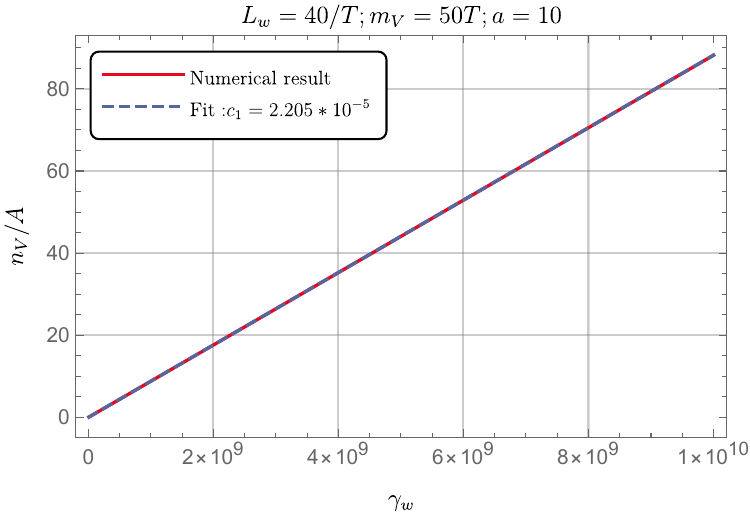}
    \includegraphics[scale=0.6]{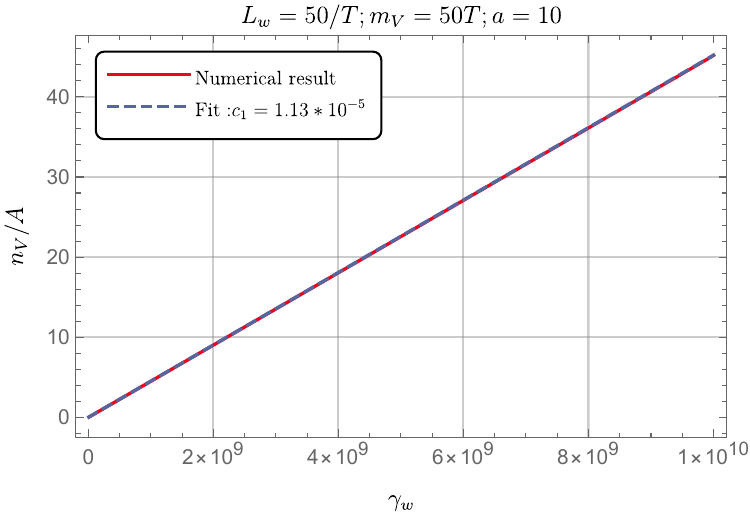}
    \caption{Comparison between the numerical results for different values of $L_w T$ and the fit using parameters given in Table~\ref{tab:fit-kappa2}.}
    \label{fig:fit-kappa2}
\end{figure}

\paragraph{Results fitting the dependence on $\bm{\kappa}$} --- Now we give the fit for the dependence of $I(\xi,\gamma_w,\kappa)$ on $\kappa$. The best fitted values of $c$ for $\kappa=10,...,50$ are given in Table~\ref{tab:fit-kappa2}. We show some examples of the comparison between the numerical results and fit function in Fig.~\ref{fig:fit-kappa2}. From the table, it is easy to deduce that 
\begin{align}
    c= 1.41\times \frac{1}{\kappa^3}\,. 
\end{align}

\begin{table*}[ht]
\centering
    \begin{tabular}{|c|c|c|c|c|c|}
    \hline
    & $\kappa=10$   & $\kappa=20$ & $\kappa=30$ & $\kappa=40$ & 50 \\
    \hline
    $c$ & $1.41\times 10^{-3}$ & $1.76\times 10^{-4}$ &  $5.22\times 10^{-5}$ & $2.205\times 10^{-5}$ & $1.13\times 10^{-5}$\\
    \hline
    \end{tabular}
    \caption{Fitted $c$ for different values of $\kappa$.}
    \label{tab:fit-kappa2}
\end{table*}

Finally, we obtain the produced density for vector dark matter in the plasma frame,
\begin{align}
    \boxed{n_V \approx 6.9\times 10^{-4}\times \frac{\gamma_w T^4 \lambda^2 v_b^2 }{L_w m_V^4}\,.}
\end{align}
Again, assuming no additional processes that decrease the dark matter density except for the redshift, one
obtains the relic abundance of the vector dark matter today,
\begin{align}
\label{eq:OmegaV}
    \boxed{\overbar{\Omega}_{V,\rm BE}^{\rm today} h^2\approx  4.1 \times \left(\frac{\gamma_w}{10^{4}}\right) \left(\frac{\lambda}{0.1}\right)^2 \left(\frac{10}{L_w T_n}\right)\left(\frac{100 T_n}{m_V}\right)^2 \left(\frac{100 v_b}{m_V}\right)\left(\frac{100}{g_{\star S} (T_{\rm after})}\right)\left(\frac{ v_b}{100 \rm GeV}\right)\left(\frac{T_n}{T_{\rm after}}\right)^3\,.}
\end{align}
Comparing with Eq.~\eqref{eq:Omegaphi}, we see that the production of vector dark matter from bubble expansion grows without bound as $\gamma_w$ increases. This can easily lead to a strong overproduction of heavy vector dark matter from bubble expansion for sufficiently large $\gamma_w$. Note that, due to potential unitarity saturation, the linear growth in $\gamma_w$ may change to a weaker growth for sufficiently large $\gamma_w$; however, this is beyond the regions of $\gamma_w$ of our phenomenological study. In the next Section, we show that the effect on bubble wall friction from vector dark matter production does not prevent reaching such relativistic velocities, before going on in Sec.~\ref{sec:vectorproduction} to discuss the subsequent dark matter evolution after being produced at the phase transition.

\section{Bubble wall friction from heavy dark matter production}
\label{sec:friction}

Any particle processes that involve interactions with the wall, either due to a background-field-dependent mass term or a background-field-dependent vertex, can generate friction on the wall. Since the particle production mechanism studied in this work assumes fast bubble walls, we must check whether or not the pair production processes under study would create too much friction on the wall to spoil this assumption. In this Section, we summarise the cases of bubble wall friction from scalar production and transition radiation and compare them with our calculation of friction due to vector boson pair production.  

Bubble wall dynamics are very complicated. The situation is simplified dramatically if one assumes that the wall motion has already entered into the so-called detonation regime and furthermore assumes that the wall velocity is large enough such that one can ignore the collisions between particles when they cross the wall (the so-called ballistic limit). We refer to the regime that satisfies these two assumptions as the {\it ballistic-detonation regime}. The analysis below adopts this simplification. However, we note that it is possible that the bubble wall never enters such a regime due to hydrodynamic obstruction, a frictional pressure barrier purely caused by inhomogeneous fluid temperature and velocity distributions across the wall~\cite{Konstandin:2010dm,Ai:2021kak,Cline:2021iff,Laurent:2022jrs,Ai:2023see,Ai:2024btx}. See in particular a recent analysis~\cite{Ai:2024shx}.\footnote{See, however, Ref.~\cite{Krajewski:2024gma} for a discussion on the stability of the hydrodynamic obstruction.}

The driving force on the wall is identified as the zero-temperature potential difference between the inside and outside of the wall~\cite{Bodeker:2009qy,Ai:2024shx} 
\begin{align}
    \P_{\rm driving}=\Delta V\,.
\end{align}
In the ballistic-detonation regime described above, the friction on the wall can be studied on a process-by-process basis. At the leading order in $\gamma_w$, the contribution is from the $1\rightarrow 1$ process~\cite{Bodeker:2009qy} 
\begin{align}
    \mathcal{P}_{1\rightarrow 1}\simeq \sum_i g_i c_i \frac{\Delta m^2_i  T^2}{24}\,,
\end{align}
where $c_i=1(1/2)$ for bosons (fermions), $g_i$ is the number of internal degrees of freedom, $T$ is the nucleation temperature, and $\Delta m_i$ is the mass gain of the particle as it transitions from the exterior to the interior of the bubble. 

At the next-leading order in $\gamma_w$, the contribution is from $1\rightarrow 2$ transition radiation, wherein a fermion hitting the wall emits a soft gauge boson~\cite{Bodeker:2017cim,Gouttenoire:2021kjv},\footnote{Ref.~\cite{Hoche:2020ysm} finds that the friction from transition radiation scales as $\gamma_w^2$, though it has been argued that this may in part be due to different assumptions in the calculation~\cite{Azatov:2020ufh,Gouttenoire:2021kjv}.}
\begin{align}
    \mathcal{P}_{1\rightarrow 2;\rm  TR} \approx C \times  \gamma_w g^2\left(\log\frac{\Delta m_{\rm g.b.}}{\mu}\right) \Delta m_{\rm g.b.} T^3\,,
\end{align}
where $g$ is the gauge coupling, $\Delta m_{\rm g.b.}\sim g v_b/\sqrt{2}$ is the mass of the gauge boson in the broken phase (in the symmetric phase, the gauge boson is assumed to be massless), $\mu$ is an IR cutoff related to the thermal mass or screening mass and roughly gives $\log(\Delta m_{\rm g.b}/\mu)\sim \O(1)-\O(10)$~\cite{Gouttenoire:2021kjv}. The numerical factor $C$ is model dependent; for example, $C\approx 1.57$ for a Higgs electroweak phase transition.

\subsection{Friction from scalar dark matter particle production \texorpdfstring{$\P_{\phi\rightarrow 2\chi}$}{TEXT}}

The friction due to the $1\to 2$ process of scalar dark matter pair production has been studied in Refs.~\cite{Azatov:2020ufh,Azatov:2021ifm,Baldes:2022oev} analytically and more precisely in Ref.~\cite{Ai:2023suz} by numerical fitting. In Ref.~\cite{Ai:2023suz}, the integral for $\P_{\phi\rightarrow 2\chi}$ has been simplified to the following form
    \begin{align}
    \label{eq:P-scalar}
    \P_{\phi\rightarrow 2\chi}&=B \times \int_{x_{\rm min}}^\infty \d x\, x^{1/2} \e^{-x}\int_0^{y_{\rm max}(x)}\d y\, y \int_{-|z|_{\rm max}(x,y)}^{|z|_{\rm max}(x,y)}\d z\, \frac{\widetilde{\Delta p^z}\, {\rm csch}^2 \left(\frac{\kappa\pi \widetilde{\Delta p^z}}{2}\right)}{\widetilde{E}\widetilde{G}^{1/2}}\,,
\end{align}
where $B=\sqrt{2\pi} \lambda^2 v_b^2 \kappa^2 T^2/( 1024\pi^2)$ and other quantities are defined in Sec.~\ref{sec:scalar-fit}. With numerical results for the integral, a fit is found to be 
\begin{align}
\label{eq:P-h-to-2phi}
    \P_{\phi\rightarrow 2\chi}\approx   0.9 \times 10^{-4}\times\lambda^2 v_b^2 T^2\log\left(1+0.26\times \frac{\gamma_w T}{L_w m_\chi^2}\right)\,,
\end{align}
with $\gamma_w$ scanned up to $10^{10}$. Again, we have corrected the result in Ref.\,\cite{Ai:2023suz} by a factor of $2\pi$.
The numerical factors may change if one scans higher and higher $\gamma_w$ in the fitting procedure. We refer the reader to Ref.~\cite{Ai:2023suz} for further details. 

\subsection{Friction from vector dark matter particle production \texorpdfstring{$\P_{\phi\rightarrow 2V}$}{TEXT}}

Here, we apply the same fitting procedure for the friction caused by heavy vector dark matter production. As for the number density, the integral for the pressure from vector dark matter pair production, $\P_{\phi\rightarrow  2V}$, can be quickly obtained by inserting the factor in Eq.~\eqref{eq:factor2} into Eq.~\eqref{eq:P-scalar}. 

\begin{figure}[H]
    \centering
    \includegraphics[scale=0.6]{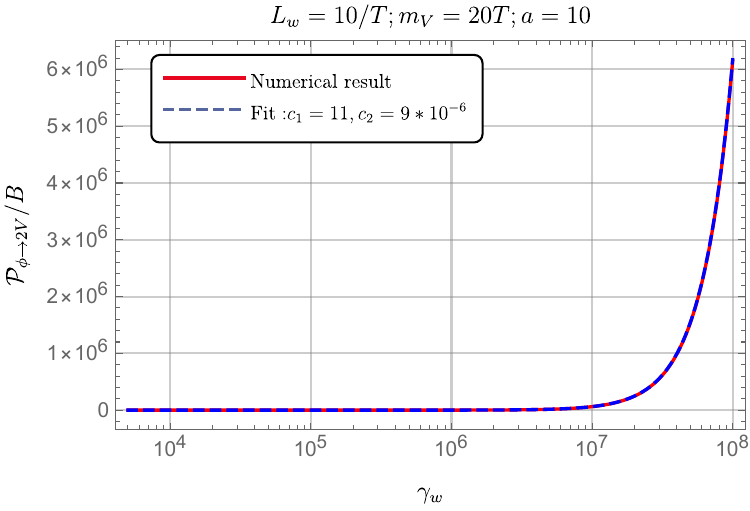}
    \includegraphics[scale=0.6]{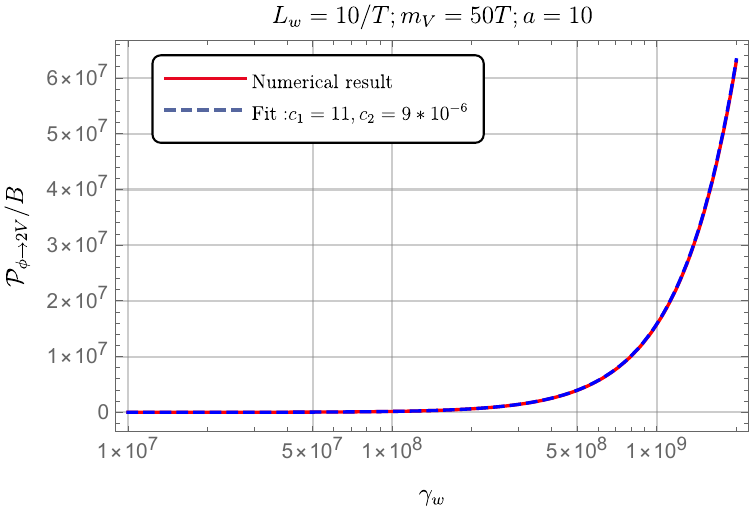}
    \includegraphics[scale=0.6]{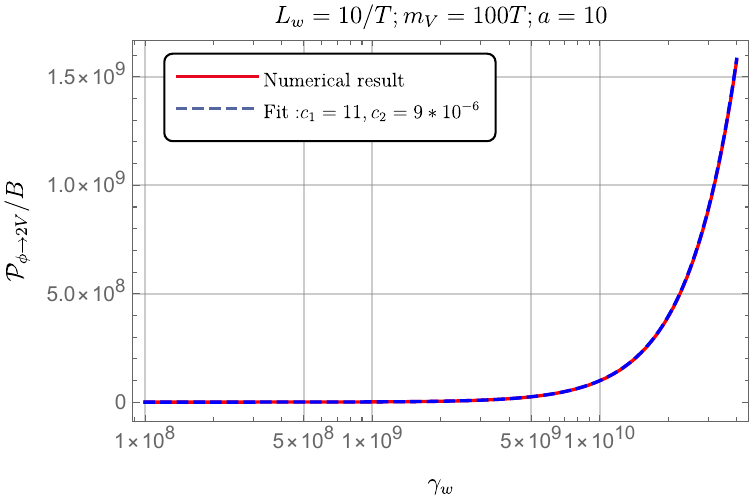}
    \caption{Plots for the numerical results of $\P_{\phi\rightarrow 2 V}/B$ vs $\gamma_w$ on a linear-log scale, for fixed $m_V/T=20, 50, 100$, respectively, and their comparison with the fit formula given in Eq.~\eqref{eq:P-fit}.}
    \label{fig:Pkappa10}
\end{figure}

\begin{figure}[ht]
    \centering
    \includegraphics[scale=0.6]{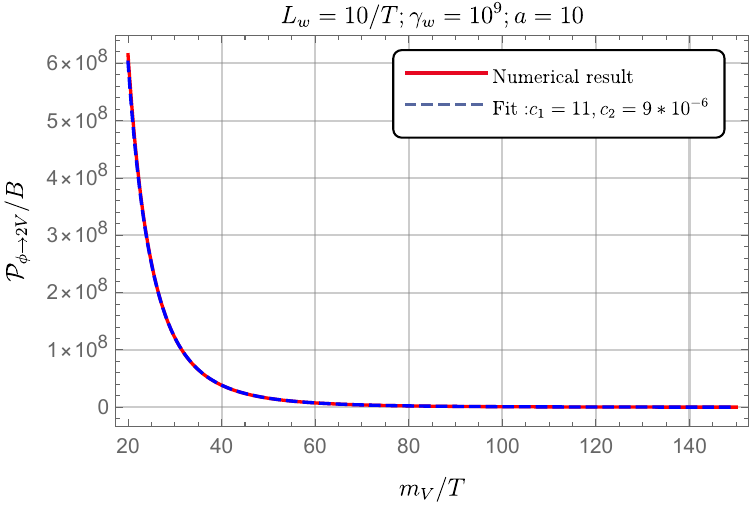}
    \caption{The numerical result of $\P_{\phi\rightarrow 2 V}/B$ as a function of $m_V/T$ for fixed $\gamma_w=10^9$, and its comparison with the fit formula.}
    \label{fig:P-fit2}
\end{figure}

\paragraph{Results for fixed $\bm{\kappa=10}$} --- When fixing $\kappa \equiv L_w T$, we find the pressure can be fitted by the following form (recall $\xi=m_V/T$)
\begin{align}
\label{eq:P-fit}
    \frac{\P_{\phi\rightarrow 2V}}{B}=c_1\times \gamma_w\times \log\left(1+c_2\times\frac{\gamma_w}{\xi^4}\right)\,.
\end{align}
For $L_w T=10$, we find $c_1=11$, $c_2=9\times 10^{-6}$. In Fig.~\ref{fig:Pkappa10}, we show $\P_{\phi\rightarrow 2V}$ as a function of $\gamma_w$ for fixed $m_V=20 T, 50 T, 100 T$, respectively. As a check, we compare the numerical result with the fit function by looking at the $\xi$-dependence in Fig.~\ref{fig:P-fit2} as well.

\begin{figure}[H]
    \centering
    \includegraphics[scale=0.6]{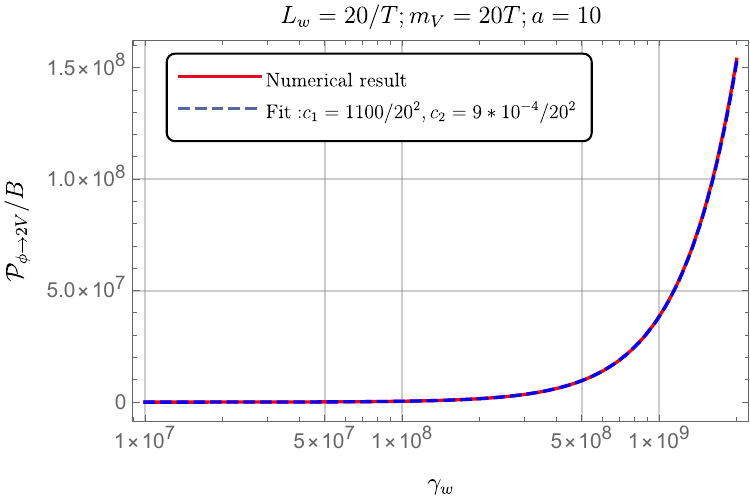}
    \includegraphics[scale=0.6]{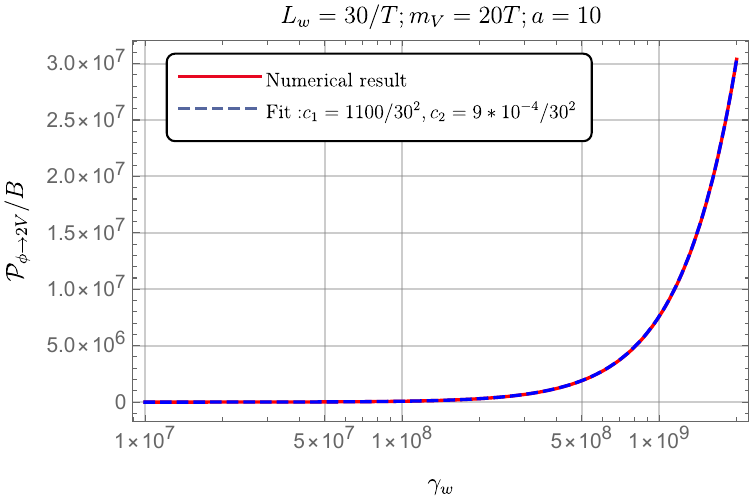}
    \includegraphics[scale=0.6]{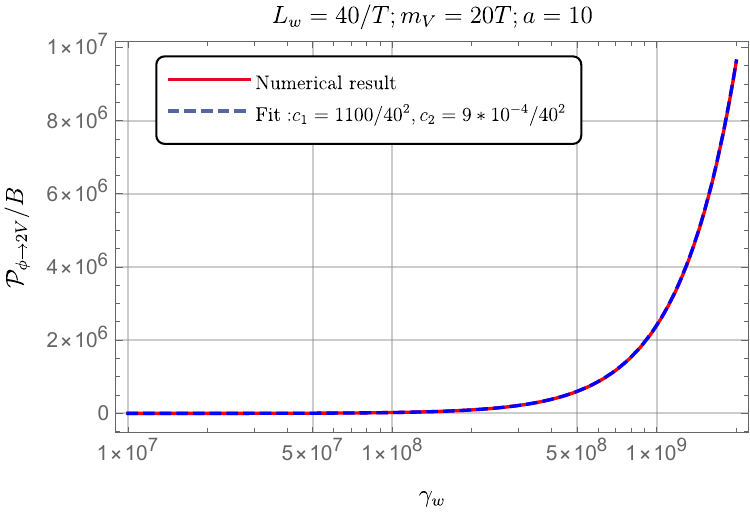}
    \includegraphics[scale=0.6]{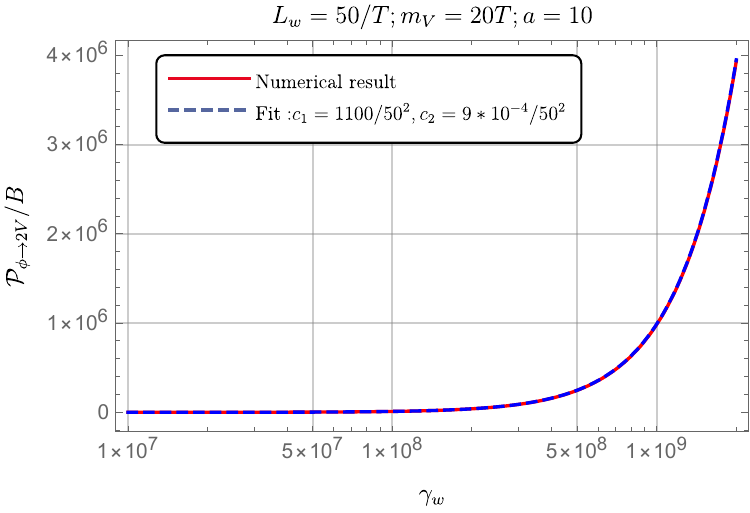}
    \caption{Comparison between the numerical results of $\P_{\phi\rightarrow 2V}/B$ for different values of $L_w T$ and the fit using parameters given in Eq.~\eqref{eq:c1c2-kappa}.}
    \label{fig:fit-kappa-friction}
\end{figure}

\paragraph{Results fitting the dependence on $\kappa$} --- Now we fit the dependence on $\kappa \equiv L_w T$ of $c_1$ and $c_2$. We find the best-fit values
\begin{align}
\label{eq:c1c2-kappa}
    c_1= 1100\times \frac{1}{\kappa^2}\,,\quad c_2=9\times 10^{-4}\times \frac{1}{\kappa^2}\,. 
\end{align}
Some examples of the comparison between the numerical results and fit function are shown in Fig.~\ref{fig:fit-kappa-friction}, and demonstrate excellent agreement.

Therefore, we finally obtain
\begin{align}
\label{eq:P_phito2V}
    \boxed{\P_{\phi\rightarrow 2V}\approx  2.7\times 10^{-1} \times \lambda^2 v_b^2 T^2\gamma_w \log\left(1+9\times 10^{-4}\times\frac{\gamma_w T^2 }{L_w^2 m_V^4}\right)\,.}
\end{align}
Note that when the second term inside the logarithm is smaller than one, 
\begin{align}
\label{eq:quadratic-scaling-regime}
  \gamma_w \lesssim 1.8\times 10^{10}\left(\frac{L_w^2 T^2}{10^2}\right)\left(\frac{m_V }{20 T}\right)^4 \qquad  {\rm (quadratic\ scaling\ regime)} \,,
\end{align}
one can Taylor-expand the logarithm and obtain
\begin{align}
\label{eq:P_phito2V-quadratic}
    \P^{(2)}_{\phi\rightarrow 2V}\approx  2.4\times 10^{-4}  \times\frac{ \lambda^2 v_b^2 \gamma_w^2 T^4 }{L_w^2 m_V^4}\,.
\end{align}
One may compare this result with that given in Ref.~\cite{Azatov:2024crd}. In Eq.~(72) of Ref.~\cite{Azatov:2024crd},  noting that $n_h\sim T^3$ and that Ref.~\cite{Azatov:2024crd} assumes $L_w\sim 1/v_b$, one gets the same dependence on $v_b$, $\gamma_w$, $T$ and $L_w$. The superscript ``$(2)$'' reminds us that this expression is only valid in the quadratic scaling regime defined by Eq.~\eqref{eq:quadratic-scaling-regime}. For larger $\gamma_w$, the quadratic scaling may be replaced by the linear-logarithmic scaling, Eq.~\eqref{eq:P_phito2V}. However, we note that the scanned range for $\gamma_w$ in the above fitting procedure does not go beyond the quadratic scaling regime; the quadratic scaling may therefore be valid for higher $\gamma_w$ and could even be exact.\footnote{The reason we chose the logarithmic function is that the same logarithmic structure for the scalar case, $\P_{\phi\rightarrow 2\chi}$, found in Ref.~\cite{Ai:2023suz} has analytic support~\cite{Baldes:2022oev}.} We leave this open question for future study. We also note that a quadratic behaviour, valid up to  $\gamma_w\sim 1/(m_V L_w)$, was previously observed for the friction $\P_{V\rightarrow V}$ caused by the 1-to-1 process of the vector boson crossing the wall, which is associated with a factor of $\rho_V (\Delta m^2_V/m^2_V)^2$ where $\rho_V$ is the energy density of the vector boson outside of the bubble~\cite{GarciaGarcia:2022yqb}. In our case, this contribution is negligible due to the suppression from both $\rho_V$ and $(\Delta m^2_V/m^2_V)^2$ for large $m_V$.

Finally, we caution that due to unitarity saturation the behaviour~\eqref{eq:P_phito2V} may not be valid for all $\gamma_w$. Although valid for the scanned regions of $\gamma_w$ (cf. Fig.~\ref{fig:dP1} and the discussion below Eq.~\eqref{eq:fit-function2}), we expect that the linear-logarithmic dependence on $\gamma_w$ of $\P_{\phi\rightarrow 2V}$ may smoothly change to the logarithmic dependence for sufficiently large $\gamma_w$. We leave a more extensive study of the friction for future work.

\subsection{Terminal wall velocity in the ballistic-detonation regime}
\label{subsec:terminal}

Transition radiation friction, with its linear dependence in $\gamma_w$, severely restricts the terminal wall velocity. Even if we assume that the wall velocity gets past the hydrodynamic obstruction and enters into the ballistic-detonation regime, a large Lorentz factor $\gamma_w$ requires a very strong supercooling in the presence of transition radiation. According to Ref.~\cite{Gouttenoire:2021kjv}, for a first-order electroweak phase transition, one has $\gamma_w\sim \alpha_n^{3/4}$.\footnote{See Eq.~(6.10) of Ref.~\cite{Gouttenoire:2021kjv} and notice that $T_{\rm start}/T_{\rm nuc}\approx \alpha_n^{1/4}$.} On the other hand, there is a maximal phase transition strength $\alpha_n$ above which the phase transition can never complete~\cite{Guth:1982pn,Ellis:2018mja}. This makes large Lorentz factors unlikely for a first-order electroweak phase transition. See also Ref.~\cite{Banerjee:2024qiu} for an analysis based on the Standard Model effective field theory where it is shown that only $\gamma_w\lsim 10$ is feasible. However, if we consider a dark sector FOPT where $\phi$ couples only with the dark matter (and perhaps also the Higgs field), then the transition-radiation process is absent and it is possible to have very large Lorentz factors $\gamma_w$. We now show that this is still the case when including friction from vector boson pair production, despite a quadratic scaling in $\gamma_w$.

Let us estimate the terminal wall velocity in the ballistic-detonation regime assuming that $\phi$ minimally couples with the dark matter vector field. In this case, we only have two leading contributions for the frictional pressure, $\P_{\phi\rightarrow \phi}$ from the 1-to-1 process, and $\P_{\phi\rightarrow 2V}$ from the 1-to-2 process.\footnote{Depending on the potential $V(\Phi)$, we may also have the process $\phi\rightarrow 2\phi$. But we expect $\P_{\phi\rightarrow 2\phi}$ to have a logarithmic dependence on $\gamma_w$ similarly to the friction from heavy scalar dark matter production, $\P_{\phi\rightarrow 2\chi}$, and $\P_{\phi\rightarrow 2\phi}\ll \P_{\phi\rightarrow\phi}$~\cite{Ai:2023suz}.  } We do not need to consider $\P_{V\rightarrow V}$ as the number density of dark matter particles outside of the wall is suppressed given $m_V\gg T_n$. The condition
\begin{align}
    \Delta V \gg P_{\phi\rightarrow\phi}
\end{align}
can easily be arranged with a properly chosen potential $V(\Phi)$.
Therefore, the terminal velocity is most sensitive to $\P_{\phi\rightarrow 2V}$. We then have the equation
\begin{align}
    \P_{\phi\rightarrow 2V} (\gamma_w; T_n) =(\Delta V-\P_{\phi\rightarrow\phi}) \approx \Delta V\,.
\end{align}
Dividing both the LHS and RHS by $T_n^4$, we get
\begin{align}
     8.2\times \left(\frac{\lambda}{0.1}\right)^2 \left(\frac{v_b}{T_n}\right)^2 \left(\frac{\gamma_w}{10^5}\right) \log \left[ 1+9\times \left(\frac{\gamma_w}{10^{14}}\right) \left(\frac{10^2}{L_w^2 T_n^2}\right)\left(\frac{100 T_n}{m_V}\right)^4\right]= \left( \frac{g_\star (T_n)}{100}\right)\left(\frac{\alpha_n}{1}\right)\,,
\end{align}
where we have used the definition for $\alpha_n$, Eq.~\eqref{eq:alpha-n}.
In the quadratic scaling regime defined by Eq.~\eqref{eq:quadratic-scaling-regime} (with $T\rightarrow T_n$), we obtain 
\begin{align}
\gamma_w \approx  4\times 10^8 \left( \frac{g_\star (T_n)}{100}\right)^{\frac{1}{2}}\left(\frac{\alpha_n}{1}\right)^{\frac{1}{2}}\left(\frac{0.1}{\lambda}\right)\left(\frac{T_n}{v_b}\right)\left(\frac{ L_w T_n}{10}\right)\left(\frac{m_V}{100 T_n}\right)^2\,.
\end{align}
We see that the bubble wall can indeed reach high terminal wall velocities. Note that the terminal Lorentz factor is sensitive to the ratio $(m_V/T_n)$ but has a weak dependence on $T_n$ itself so that one can consider a dark sector FOPT almost at any energy scale to get large terminal Lorentz factors.

\section{Parameter space of heavy WIMP vector dark matter}
\label{sec:wash-out}

Having shown that ultra-relativistic bubble expansion during a FOPT is both possible and can efficiently source a number density of heavy vector dark matter, we now consider the implications of this new production mechanism on the viability of TeV-scale WIMP vector dark matter. Specifically, we focus on the region of parameter space where thermal freeze-out is unable to generate the observed relic abundance. For non-thermal production mechanisms of vector dark matter to dominate over thermal freeze-out, we require its scalar coupling to be sufficiently large for the thermal relic density to be negligible. Non-thermal bubble wall production at the time of the phase transition could then generate a greater abundance than the observed relic density. However, its subsequent Boltzmann evolution must be taken into account which can reduce the non-thermally produced dark matter to its observed relic abundance. Planck constraints on $\Omega_{\rm{M}0}h^2$ and $\Omega_{{\rm B}0}h^2$ tell us that $\Omega_{{\rm DM}0}h^2=0.121$ \cite{Planck:2018vyg} is a good approximation to a couple of percent, which is a good enough approximation for this work.  The subscripts ``M'', ``DM'' and ``B'' correspond to matter, dark matter and baryons, respectively, and as usual the subscript $0$ refers to quantities measured at $t=t_0$, (i.e. today, at redshift $z=0$). The dark matter relic abundance is
\begin{equation}
\Omega_{{\rm DM}0}h^2=0.121=\frac{\rho_{{\rm DM}0}}{\rho_{c}}h^2=\frac{8\pi G\rho_{{\rm DM}0}h^2}{3 H_0^2}=\frac{8\pi G\rho_{{\rm DM}0}}{3\zeta^2}
\end{equation}
where $\zeta=100$ km s$^{-1}$Mpc$^{-1}=2.133\times 10^{-42}$ GeV. 
 The density of dark matter is then given by 
\begin{equation}
\rho_{{\rm DM}0}=Y_{{\rm DM}0}s_0E_{{\rm DM}0}=8.014\times 10^{-47}\rm{GeV}^4
\end{equation}
where $Y=n/s$ where $n$ is number density and $s$ is entropy density. If the kinetic energy of dark matter is negligible, which we know today it is to a good approximation, $E_{{\rm DM}0}=m_{\rm DM}$. 

Let us first consider the thermal freeze-out scenario before treating the non-thermal case with simplified Boltzmann equations, following Ref.~\cite{Falkowski:2012fb}. For temperatures much lower than the mass of the vector boson, $m_V \gg T$, the thermally averaged cross-section of dark matter annihilation through its scalar coupling is approximately given by
\begin{equation}
\left. \langle \sigma v \rangle \right|_{m_V \gg T} \simeq \frac{\lambda^2}{64 \pi m_V^2} \, .
\end{equation}
We assume here only that the scalar mediates annihilation to lighter states through model-dependent couplings that we leave unspecified as we are agnostic about the scalar interactions to the visible sector. 
This cross-section fixes a lower bound on the coupling to avoid over-producing dark matter when annihilations are no longer sufficiently efficient for cross-sections below $\langle \sigma v \rangle_\text{thermal relic} \simeq 1.9 \times 10^{-9} \text{ GeV}^{-2}$~\cite{Cirelli:2024ssz}, which sets the thermal relic density to the observed value. We may write this thermal over-production bound, for $m_V \gg T$, as 
\begin{equation}
    \lambda \gtrsim 0.6 \left( \frac{m_V}{10^3 \text{ GeV}} \right) \, .
    \label{eq:thermalrelicbound}
\end{equation}

Now, for couplings greater than Eq.~\eqref{eq:thermalrelicbound}, the thermal relic density will be a fraction of dark matter and the remaining dark matter abundance can be due to non-thermal production from bubble walls. We assume the non-thermal production to be strong enough to produce a greater dark matter abundance than the observed relic density. This can easily be the case for high bubble wall velocities with sufficiently large $\gamma_w$ factors, as shown in the previous Section. We also assume the phase transition to occur at a temperature below the freeze-out temperature,\footnote{Here we do not consider exceptionally strong supercooling such that we have $T_n\approx T_{\rm after}$ and in this Section we refer to either of them as the phase transition temperature $T_{\rm PT}$.} $T_{\rm PT}<T_{\rm FO} \simeq m_V/20$. If $T_{\rm PT}>T_{\rm FO}$, then the produced dark matter particles would reach thermal equilibrium again and one simply obtains the standard thermal freeze-out relic abundance. Unlike the thermal case where annihilations are no longer efficient below $T_{\rm FO}$, the non-thermal abundance can still be partially washed out for $T < T_{\rm PT}$, as we will now see.

We start with the integrated Boltzmann equation
\begin{align}
\label{eq:boltzman-eq1}
    a^{-3} \frac{\d (n_V a^3)}{\d t}=\langle \sigma v\rangle \left[\left(n_V^{\rm eq}\right)^2-n_V^2\right]\,.
\end{align}
In terms of the co-moving number density $Y\equiv n_V/s$ where $s=2\pi^2 g_{\star S}(T) T^3/45$ is the entropy density, and the dimensionless time variable $x\equiv m_V/T$, one can write Eq.~\eqref{eq:boltzman-eq1} as
\begin{align}
    \frac{\d Y}{\d x}=-\frac{\alpha}{x^2} \left[Y(x)^2-Y_{\rm eq}(x)^2\right]\,,
\end{align}
where 
\begin{align}
    \alpha\equiv\frac{2\pi^2}{45} g_{\star S}(T)\sqrt{\frac{90}{8\pi^3 g_{\star}(m_V)}} M_{\rm Pl}m_V \langle \sigma v\rangle\,,
\end{align}
with $g_\star (m_V)$ being the number of relativistic degrees of freedom evaluated at $T=m_V$. Above, we have used $\d x/\d t= H x$ (recall $T\propto a^{-1}$), the relation $H(T)=H(m_V)/ x^2$ that is valid for a radiation-dominated universe, and also the Friedmann equation~\eqref{eq:Friedmann}. Below, we ignore the temperature-dependence in $g_{\star S}$ and $g_\star$ and take $g_{\star S}\approx g_\star\approx 100$.

Assuming $Y(x) \gg Y_\text{eq}(x)$, one simply gets
\begin{equation}
    \frac{\d Y}{\d x} = -\frac{\alpha}{x^2} Y(x)^2 \,. 
    \label{eq:boltzmann}
\end{equation}
Integrating Eq.~\eqref{eq:boltzmann} from the time of the phase transition at $x_{\rm PT}$ to later times at lower temperatures, $x \gg x_{\rm PT}$, gives 
\begin{equation}
    \frac{1}{\alpha}\left(\frac{1}{Y(x)} - \frac{1}{Y(x_{\rm PT})}\right) = \frac{1}{x_{\rm PT}} - \frac{1}{x} \, . 
\end{equation}
Since $x_{\rm PT} \ll x$ and the non-thermal dark matter density produced at the phase transition is much larger than the observed relic density at $x \to \infty$, $Y(\infty) \ll Y(x_{\rm PT})$, this simplifies to
\begin{equation}
    Y(\infty) \simeq \frac{x_{\rm PT}}{\alpha}  \, .
    \label{eq:nonthermaldensity}
\end{equation}
Fixing the temperature of the phase transition and the annihilation cross-section, through the coupling and mass of the vector dark matter, then determines the resulting relic density at late times. We may compare the necessary cross-section for setting this non-thermal dark matter relic density with the cross-section that would give a thermal relic abundance in agreement with the CMB. The thermal relic density from freeze-out can be approximated by
\begin{equation}
    Y_\text{thermal}(\infty) \simeq \frac{x_{\rm FO}}{\alpha_\text{thermal}} \, ,
    \label{eq:thermaldensity}
\end{equation}
where $\alpha_\text{thermal} \equiv \frac{2\pi^2\sqrt{g_\star}}{45}\times \sqrt{90/8\pi^3} M_{\rm Pl} m_V \langle \sigma v \rangle_\text{thermal relic}$ and $x_{\rm FO} = m_V / T_{\rm FO}$. Comparing Eqs.~\eqref{eq:nonthermaldensity} and~\eqref{eq:thermaldensity}, we see that the cross-section required for non-thermal production to give the observed dark matter relic density is greater than the one that would result from the thermal freeze-out scenario, 
\begin{equation}
    \langle \sigma v \rangle = \frac{T_{\rm FO}}{T_{\rm PT}} \langle \sigma v \rangle_{\rm thermal\ relic} \, .
\end{equation}
The value of $\lambda$ that ensures the correct non-thermal relic abundance at late times can then be written as
\begin{equation}
    \lambda \simeq 0.4 \left(\frac{m_V}{10^3\text{ GeV}}\right)^{3/2} \left(\frac{10^2\text{ GeV}}{T_{\rm PT}}\right)^{1/2} \, .
    \label{eq:nonthermallambda}
\end{equation}
The higher cross-section for vector dark matter from bubble walls relative to the thermal freeze-out case may improve its prospects of direct and indirect detection. 
These limits are not shown as this will depend on the model-dependent scalar coupling to the visible sector that we have not specified.  

\begin{figure}[!t]
\centering
\includegraphics[scale=0.8]{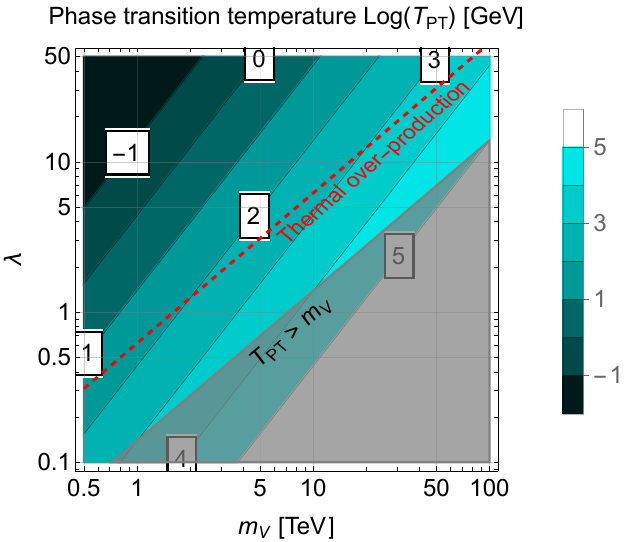}
\includegraphics[scale=0.8]{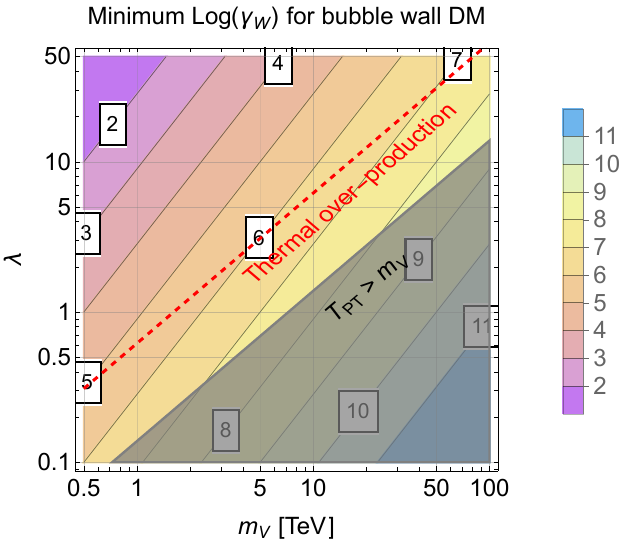}
\caption{Left: Log contours of the phase transition temperature in GeV in the $(m_V, \lambda)$ plane where the observed vector dark matter density is obtained assuming an initial non-thermal over-production. Right: Log contours of the minimum $\gamma_w$ needed for non-thermal production of vector dark matter from bubble wall expansion to exceed the observed relic density. In both plots, the region below the dashed red line is excluded by over-production of vector bosons through thermal freeze-out if the inflationary reheating temperature was sufficiently high, and the grey region is excluded by the required phase transition temperature being larger than the vector boson mass. }
\label{fig:nonthermalabundance}
\end{figure}

The left plot of Fig.~\ref{fig:nonthermalabundance} shows contours of the phase transition temperature, in units of GeV on a logarithmic scale, for the parameter space of $m_V$ vs $\lambda$ with the observed dark matter relic abundance set by Eq.~\eqref{eq:nonthermallambda}. The over-production of vector dark matter from thermal freeze-out corresponds to the region below the dashed red line; this region is excluded if the reheating temperature prior to the phase transition is sufficiently high for the vector boson to have been in thermal equilibrium. The grey region is where the temperature of the phase transition is larger than the mass of the vector boson. We see that the phase transition temperature is restricted to be in the sub-GeV to $\mathcal{O}(100)$ GeV range and can be at most $\mathcal{O}(10)$ TeV. 

It may be tempting to identify the phase transition with the electroweak phase transition and the scalar $\phi$ with the Higgs boson. Unfortunately, as discussed in the previous Section, transition radiation that involves a scalar or fermion radiating off a vector boson when passing through the bubble wall prevents it from attaining the large $\gamma_w$ factors necessary for our particle production to be efficient. The minimum value of $\gamma_w$ for non-thermal particle production to exceed the observed relic density of $\Omega_{{\rm DM}0}h^2=0.121$ \cite{Planck:2018vyg} is shown in logarithmic contours on the right plot of Fig.~\ref{fig:nonthermalabundance}. Since $\gamma_w \propto \alpha_n^{3/4} \lesssim \mathcal{O}(1)$ for a Higgs bubble wall in the Standard Model, this points towards phase transitions in dark sectors without transition radiation such that the bubble wall can attain much higher relativistic velocities. We give a detailed discussion on the terminal wall velocity in Section~\ref{subsec:terminal}.  

We have focused on the WIMP vector dark matter scenario where the coupling is $\mathcal{O}(1)$ or larger and determines the annihilation cross-section for setting the relic density. We conclude that non-thermal production from bubble wall expansion can set the correct WIMP vector dark matter relic abundance in regimes where thermal production is inefficient and the non-thermal abundance generated from bubble expansion is initially over-produced. For a given phase transition temperature, the scalar coupling and mass of the vector boson are set by the requirement to annihilate the initial non-thermal vector boson density down to the observed DM relic density, in a way that is insensitive to the initial abundance. 

The parameter space could open up if one considers superheavy dark matter, such as WIMP-zillas~\cite{Chung:1998ua,Chung:1998zb,Kolb:1998ki}, or feebly interacting massive particles (FIMPs)~\cite{McDonald:2001vt,Hall:2009bx} with very small couplings. The thermal relic bound can be evaded if the corresponding freeze-out temperature, $T_{\rm FO}$, is even larger than the maximal reheating temperature of the universe. In this case, one needs to compare the dark matter relic abundance generated from bubble expansion with that from freeze-in. This scenario for vector dark matter has recently been studied in Ref.~\cite{Azatov:2024crd}.

\section{Gravitational wave signals}
\label{sec:GW}

The previous analysis is generic because it does not specify the model, i.e., $V(\Phi)$, for the phase transition itself, leaving a large scope for model building. For WIMP vector dark matter, the most important constraint we obtain is that the phase transition temperature is restricted in the range from sub-GeV to $\O(10)$ {\rm TeV}. Together with the requirement that $\gamma_w$ is extremely large, the mechanism studied in this work may have unique features in stochastic GW signals from the FOPT.

GW signals from FOPTs depend on four key quantities, $(H,\beta,\alpha_n,v_w)$. Usually, these quantities are computed at the percolation time (with a corresponding temperature usually denoted by $T_*$) which is later than the nucleation time. For moderately strong phase transitions, $T_n$ and $T_*$ are very close to each other and one usually does not distinguish between them. We thus use a unique phase transition temperature $T_{\rm PT}$ as we did in the last Section. For the GW signals, only the dimensionless combination $\beta/H$ (estimated at $T_{\rm PT}$) matters. We shall consider two typical values of it: $\beta/H=100$ and $\beta/H=1000$, while fixing $\alpha_n=1$. 

Following Ref.~\cite{Azatov:2024crd}, we use the bulk flow model~\cite{Konstandin:2017sat} for the GW power spectrum for bubble expansion with large Lorentz factors. Simply taking $v_w\rightarrow 1$, the power spectrum reads~\cite{Konstandin:2017sat}
\begin{align}
    \Omega_{\rm GW} h^2=\Omega_{\rm peak} h^2\times S(f,f_{\rm peak})\,,
\end{align}
where 
\begin{align}
    S(f,f_{\rm peak})=\frac{(a+b) f_{\rm peak}^b f^a}{b f_{\rm peak}^{a+b}+a f^{a+b}}\,,\qquad (a= 0.9,b= 2.1)
\end{align}
and
\begin{subequations}
\begin{align}
&\Omega_{\rm peak}h^2\approx 1.07 \times 10^{-6}\left(\frac{H}{\beta}\right)^2 \left(\frac{\alpha_n \kappa_f}{1+\alpha_n}\right)^2\left(\frac{100}{g_\star(T_{\rm PT})}\right)^{1/3}  \,,\\
& f_{\rm peak}\approx 2.12\times 10^{-3}\, {\rm  mHz}\left(\frac{\beta}{H}\right)\left(\frac{T_{\rm PT}}{100 {\rm GeV} }\right)\left(\frac{g_\star(T_{\rm PT})}{100}\right)^{1/6}\,. \label{eq:fpeak}
\end{align}
\end{subequations}
Above, $\kappa_f$ is an efficiency factor quantifying the fraction of the available vacuum energy that goes into the kinetic energy of the fluid. We shall take $\kappa_f\simeq 1$ for simplicity.

\begin{figure}[t!]
    \centering
    \includegraphics{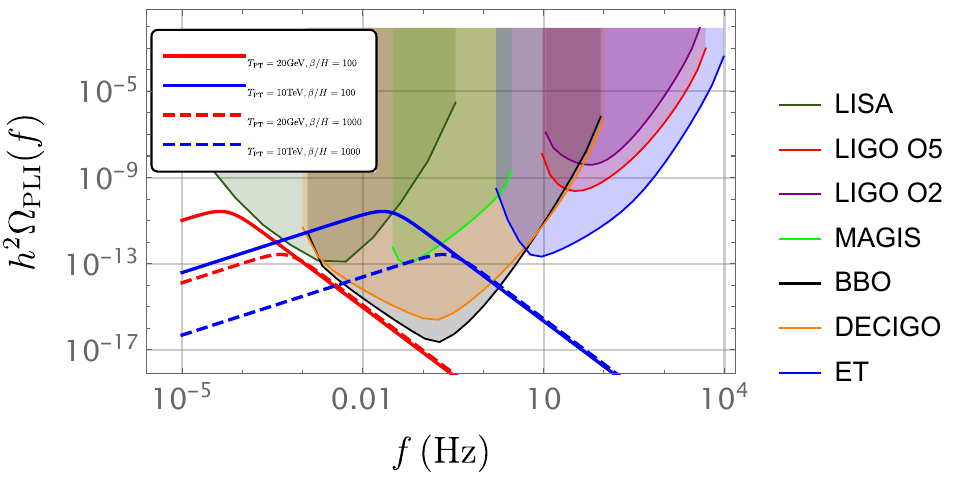}
    \caption{GW signals predicted by the bulk flow model~\cite{Konstandin:2017sat} for fixed $\alpha_n=1$. The dark matter production mechanism studied in this work could be associated with GW signals that are within reach of future GW experiments. The PLI sensitivity curves are generated based on Ref.~\cite{Azatov:2019png}, with strain noise extracted from Refs.~\cite{KAGRA:2013rdx,LIGOScientific:2014pky,Robson:2018ifk,Graham:2017pmn,Yagi:2011yu,Yagi:2013du,Sathyaprakash:2012jk}.}
    \label{fig:GWs}
\end{figure}

In Fig.~\ref{fig:GWs} we show the GW signals predicted by the bulk flow model for the parameter space region relevant for the dark matter production, compared to the power-law integrated (PLI) sensitivity curves of various current and future experiments. For simplicity, we have fixed $\alpha_n$ to be one. A smaller $\alpha_n$ would reduce the amplitude of the GW power spectrum. We can see that the phase transition temperature $T_{\rm PT}=20 {\rm GeV}$ (and $\alpha_n=1, \kappa=1$), denoted by the solid and dashed red lines for $\beta/H$ values of 100 and 1000 respectively, is near the edge of the PLI sensitivity curves of LISA, BBO, and DECIGO. For higher temperatures, the predicted GW signals could be within reach of more future GW experiments such as MAGIS and the Einstein Telescope. This will be subject to foreground uncertainties involving astrophysical processes.

\section{Conclusion}
\label{sec:Conc}

Cosmological FOPTs are motivated by a wide variety of beyond the Standard Model extensions and could provide new non-thermal mechanisms for dark matter production. Dark matter much heavier than the phase transition scale ($m_{\rm DM} > 20 T_{\rm PT}$), which would have been out of equilibrium and suppressed right before the phase transition, can be produced from bubble expansion and collision in the case of ultra-fast walls. The production of heavy scalar, fermion, and vector dark matter from bubble collisions has been studied in Refs.~\cite{Falkowski:2012fb, Giudice:2024tcp}, while scalar dark matter production has been investigated in Ref.~\cite{Azatov:2021ifm}. However, the production of heavy vector dark matter from bubble expansion has not been thoroughly explored. The effect of vector boson pair production on the bubble wall dynamics is also an interesting open question, given that transition radiation involving the emission of a single vector boson introduces a friction linear in the wall boost factor $\gamma_w$ that prevents the wall from reaching fast terminal velocities.    

In this work, we have carefully studied the production of vector dark matter pairs from bubble expansion for walls with very large Lorentz boost factors. We numerically computed the vector dark matter density and provided an empirical analytical formula fit to the numerical calculation. Our results show that the vector dark matter density scales with $\gamma_w$, differing from the scalar dark matter case~\cite{Azatov:2021ifm}. Consequently, vector dark matter can be easily produced during the FOPT. 

This mechanism requires very large Lorentz factors $\gamma_w$ for the bubble wall. For an electroweak FOPT, this is unlikely due to transition radiation friction unless there is exceptional supercooling. Therefore, we considered a dark sector phase transition. We computed the friction generated from the $\phi\rightarrow \phi$ process ($\P_{\phi\rightarrow\phi}$) and the $\phi\rightarrow 2V$ process ($\P_{\phi \rightarrow 2V}$). We found that these pressures allow for large $\gamma_w$, making the mechanism viable. Additionally, we discovered that $\P_{\phi\rightarrow 2V}$ scales as $\gamma_w\log(1+c \gamma_w)$ or $\gamma_w^2$, which differs from the scaling of transition radiation. To the best of our knowledge, this new scaling behaviour for vector boson pair production has not yet been noticed in the literature. Although our computation assumed $m_V\gg T_{\rm PT}$, we expect this scaling to persist for smaller $m_V$ and potentially be valid for the processes $h\rightarrow W^+ W^-$ and $h\rightarrow ZZ$ in an electroweak FOPT. We also gave a warning that unitarity might change this new scaling behaviour for sufficiently large $\gamma_w$ beyond the regime studied here. A check for this would be to consider a UV completion of the theory given in Eq.~\eqref{eq:Lagrangian-Proca}. We leave these interesting questions for future work.

We then studied the subsequent evolution of the non-thermally produced WIMP vector dark matter with masses at the TeV scale, assuming an initial over-production in the phase transition then accounting for dark matter annihilation in their Boltzmann equations and found that the final relic abundance can match the observed dark matter density in a wide region of mass-coupling parameter space in a way that is insensitive to the initial condition generated at the phase transition when over-production occurs. Our findings indicate that TeV-scale WIMP vector dark matter can be efficiently produced through bubble expansion in regimes where thermal freeze-out is insufficient. The phase transition temperature of the dark sector lies in a promising observational range below $\mathcal{O}(10)$ TeV where the associated stochastic GW signal may well be within reach of future GW observatories. 

Future directions to investigate include studying the impact on the electroweak FOPT and looking at the wider phenomenology of vector dark matter produced by bubble walls, both lighter and heavier than the TeV scale considered here. In this work we used an effective vector-Higgs portal parameterisation for simplicity; a more thorough study of vector dark matter production may be sensitive to details of the UV completion if the bubble wall particle production energy reaches sufficiently high scales. The direct and indirect detection prospects in dark matter searches and astrophysical signals may also provide an additional handle on the model, though this will be model-dependent on the visible sector couplings of the scalar. We leave all these interesting aspects for future work.

\section*{Acknowledgments}
 
It is a pleasure to thank Xander Nagels and Miguel Vanvlasselaer for helpful discussions and comments, especially for sharing their experience in making the GW signal sensitivity plot. We also thank Ke-Pan Xie for his comments on the manuscript. The work of WYA is supported by the UK Engineering and Physical Sciences Research Council (EPSRC), under Research Grant No. EP/V002821/1. TY is supported by United Kingdom Science and Technologies Facilities Council (STFC) grant ST/X000753/1. TY is partially supported by a Branco Weiss Society in Science Fellowship.  MF is supported in part by United Kingdom STFC Grants ST/P000258/1 and ST/T000759/1. KM is supported by an Ernest Rutherford Fellowship from the  STFC, Grant No. ST/X004155/1 and partly by the STFC Grant No. ST/X000583/1.

\newpage

\begin{appendix}
\renewcommand{\theequation}{\Alph{section}\arabic{equation}}
\setcounter{equation}{0}

 \end{appendix}

\newpage
\bibliographystyle{utphys}
\bibliography{ref}{}

\end{document}